\documentstyle{ar_prepr}
\input psfig.sty
\def\ref{\par\noindent\hangindent=1truecm}
\font\piedi=cmr8

\catcode`\@=11
\def\gsim{\ifmmode{\mathrel{\mathpalette\@versim>}}
    \else{$\mathrel{\mathpalette\@versim>}$}\fi}
\def\lsim{\ifmmode{\mathrel{\mathpalette\@versim<}}
    \else{$\mathrel{\mathpalette\@versim<}$}\fi}
\def\@versim#1#2{\lower 2.9truept \vbox{\baselineskip 0pt \lineskip 
    0.5truept \ialign{$\m@th#1\hfil##\hfil$\crcr#2\crcr\sim\crcr}}}
\catcode`\@=12
\def\pn{\par\noindent}
\def\pb{\par\noindent$\bullet\quad$}
\def\log{{\rm log\,}}
\def\lsun{\hbox{$L_\odot$}}
\def\zsun{\hbox{$Z_\odot$}}
\def\lb{\hbox{$L_{\rm B}$}}
\def\lv{\hbox{$L_{\rm V}$}}
\def\lk{\hbox{$L_{\rm K}$}}
\def\tf{\hbox{$t_{\rm F}$}}
\def\msy{\hbox{$\msun{\rm yr}^{-1}$}}
\def\l0g{\hbox{\rm log}}
\def\msun{\hbox{$M_\odot$}}
\def\mgt{\hbox{\rm Mg$_2$}}
\def\mgs{\hbox{\rm Mg$_2-\sigma$}}
\def\mgb{\hbox{{\rm Mg}$b$}}
\def\fe{\hbox{$<\!{\rm Fe}\!>$}}
\def\hbe{\hbox{${\rm H}\beta$}}

\def\mto{\hbox{$M_{\rm TO}$}}

\def\mpcc{\hbox{${\rm Mpc}^{-3}$}}

\def\zf{\hbox{$z_{\rm F}$}}

\def\ho{\hbox{$H_\circ$}}
\def\h50{\hbox{$\ho /50$}}

\def\mto{\hbox{$M_{\rm TO}$}}

\def\kms{\hbox{km s$^{-1}$}}

\def\zf{\hbox{$z_{\rm F}$}}

\def\ho{\hbox{$H_\circ$}}
\def\h50{\hbox{$\ho /50$}}
\def\re{\hbox{$R_{\rm e}$}}
\def\ie{\hbox{$I_{\rm e}$}}
\def\ku{\hbox{$\kappa_1$}}
\def\kd{\hbox{$\kappa_2$}}
\def\kt{\hbox{$\kappa_3$}}

\def\apjs{\hbox{ApJS$\;$}}
\def\apj{\hbox{ApJ$\;$}}

\def\araa{\hbox{ARA\&A$\;$}}
\def\mn{\hbox{MNRAS$\;$}}
\def\aj{\hbox{AJ$\;$}}
\def\aap{\hbox{A\&A$\;$}}

\def\ApJ{\hbox{ApJ$\;$}}
\def\nat{\hbox{Nature$\;$}}
\def\acs{\hbox{ASP Conf. Ser.$\;$}}
\def\pasp{\hbox{PASP $\;$}}

\begin{document}
\title{{STELLAR POPULATION DIAGNOSTICS OF ELLIPTICAL
         GALAXY FORMATION}}
\markboth{Alvio Renzini}{Elliptical Galaxy Formation}
\author{Alvio Renzini
\affiliation{INAF, Osservatorio Astronomico di Padova}}
\begin{abstract}

Major progress has been achieved in recent years in mapping the properties of 
passively-evolving, early-type galaxies (ETG) from the local universe
all the way to redshift $\sim 2$.  Here, age and metallicity estimates
for local cluster and field ETGs are reviewed as based on
color-magnitude, color-$\sigma$, and fundamental plane relations, as
well as on spectral-line indices diagnostics.  The results of applying
the same tools at high redshifts are then discussed, and their
consistency with the low-redshift results is assessed. Most low- as
well as high-redshift ($z\sim 1$) observations consistently indicate 1)
a formation redshift $z\gsim 3$ for the bulk of stars in cluster ETGs,
with their counterparts in low-density environments being on average
$\sim 1-2$ Gyr younger, i.e., formed at $z\gsim 1.5-2$; 2) the duration
of the major star formation phase anticorrelates with galaxy mass, and
the oldest stellar populations are found in the most massive
galaxies. With increasing redshift there is evidence for a decrease in the
number density of ETGs, especially of the less massive ones, whereas
existing data appear to suggest that most of the most-massive ETGs
were already fully assembled at $z\sim 1$. Beyond this redshift, the
space density of ETGs starts dropping significantly, and as ETGs
disappear, a population of massive, strongly clustered, starburst
galaxies progressively becomes more and more prominent, which makes
them the likely progenitors to ETGs.

\end{abstract}

\maketitle
\section{INTRODUCTION}
Following Hubble (1936), we still classify galaxies as ellipticals,
spirals and irregulars (see Sandage 2005, and references
therein). This was an eyeball, purely morphological scheme; however,
morphology correlates with the stellar population content of these
galaxies, with typical ellipticals being redder than the others, and
showing purely stellar absorption-line spectra with no or very weak
nebular emissions.  As a consequence, one often refers to early-type
galaxies (ETG), even if they are color (or spectral type) selected
rather than morphologically selected.  Furthermore, the bulges of
spirals of the earlier types show morphological as well as spectral
similarities with ellipticals, and one often includes both ellipticals
and bulges under the category of galactic spheroids.

Morphologically-selected and color-, or spectrum-selected samples
do not fully overlap. For example, in a recent
study of local ($z\sim 0$) ETGs from the Sloan Digital
Sky Survey (SDSS) all three criteria were adopted (Bernardi et
al. 2006), and the result is reported in Table 1 (M. Bernardi, private
communication).\footnote{Here, as well as through the whole paper,
the definition of the specific criteria adopted by the various authors
for their sample selections can be found in the original articles.}
From a sample including $\sim$123,000 galaxies with $14.5<r_{\rm
Petrosian}<17.77$ and $0.004<z<0.08$, ETGs have been selected in
turn by each of the three criteria (MOR, COL, SPE), and the resulting
numbers of galaxies fulfilling each of them is given on the diagonal
of the matrix. Out of the diagonal are the fractions of galaxies that
satisfy two of the criteria, as labelled in the corresponding row and
column.  So, out of the morphologically-selected ETGs, $70\%$ satisfy
also the color selection, etc. The correlation between color and
morphology selection persists at high redshift, e.g., at $z\sim 0.7$
about $85\%$ of color-selected, red-sequence galaxies are also
morphologically early-type, i.e., E/S0/Sa (Bell et al 2004a), an
estimate broadly consistent with the local one, when allowing for the
wider morphological criterion.

\begin{figure}[t]
%\vskip-0.5truecm
\centerline{\psfig{file=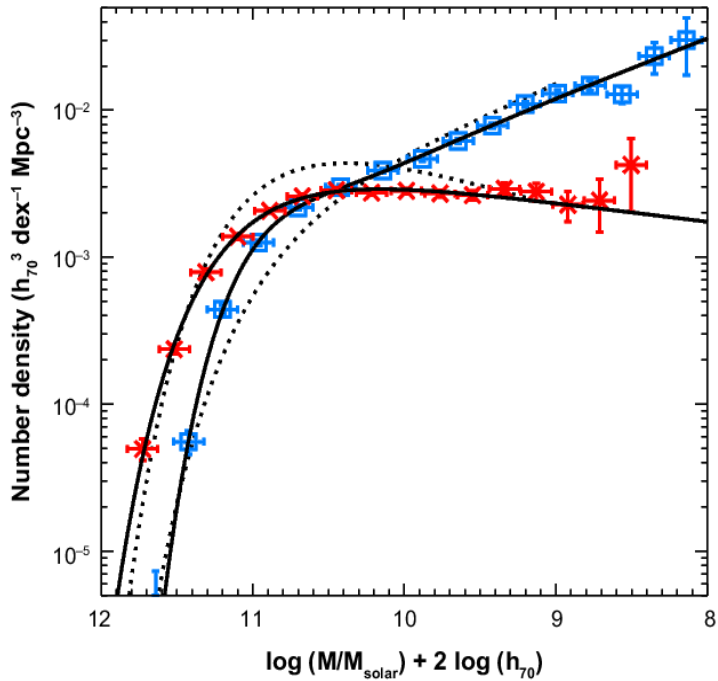,width=9.5cm,angle=0,height=9.cm}}

%\vskip-5truecm
%\plotone{feab.eps}
%\vskip 1truecm
\caption{\piedi The mass function of local ($z\sim 0$) early-type
(red) galaxies and late-type (blue) galaxies from the Sloan Digital Sky Survey
(Baldry et al. 2004). The solid lines represent best fit Schechter functions.
For the blue galaxies the sum of two different Schechter functions is
required to provide a good fit. Dotted lines show the mass functions from 
Bell et al. (2003).
}
%\vskip-5truemm
\end{figure}

\begin{figure}[t]
%\vskip-0.05truecm
\centerline{\psfig{file=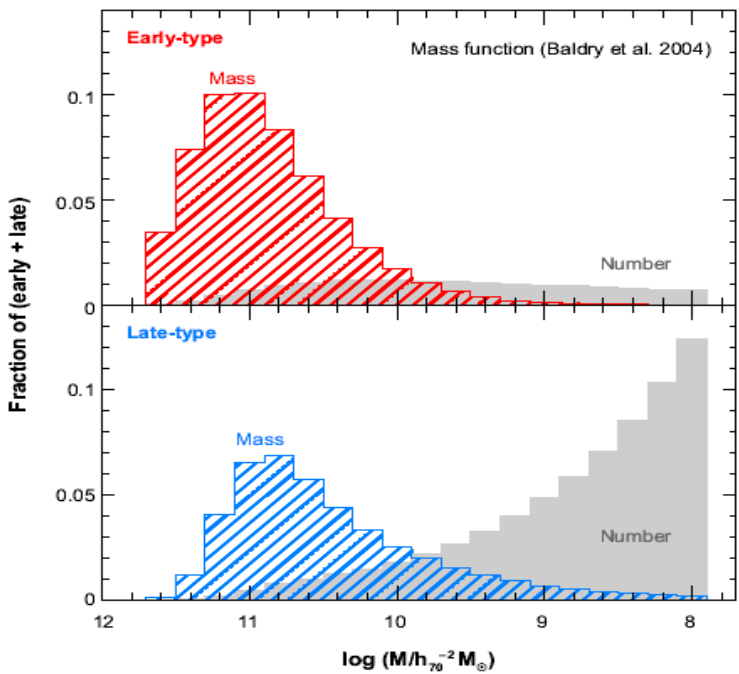,width=10cm,angle=0,height=9cm}}

\vskip 1truecm
%\plotone{feab.eps}
\vskip-1truecm
\caption{\piedi The contributions to the total stellar mass and to the
number of galaxies by early-type (red) and late-type (blue) galaxies
in the various mass bins, as derived from the best-fit mass functions
of Baldry et al. (2004) shown in Figure 1. The relative areas are
proportional to the contributions of the early- and late-type galaxies
to the total stellar mass and to the number of galaxies.  (Courtesy of
D. Thomas).  }
\end{figure}

\medskip
\begin{center}
{\bf Table 1.} Morphology- Versus Color- Versus Spectrum-Selected Samples
\end{center}
\begin{center}
\begin{tabular}{|l|l|l|l|}
\hline
&  MOR & COL  & SPE \\ \hline
MOR & 37151    &  70\%     &  81\%   \\
COL& 58\% & 44618 & 87\% \\ 
SPE & 55\% & 70\%  & 55134  \\ \hline
\end{tabular}
\end{center}

\smallskip

It is estimated that true ellipticals represent $\sim 22\%$ of the
total mass in stars in the local universe, a fraction amounting to
$\sim 75\%$ for spheroids (i.e., when including E0's and spiral
bulges), whereas disks contribute only $\sim 25\%$ and dwarfs an
irrelevant fraction (Fukugita, Hogan, \& Peebles 1998). Although earlier
estimates gave slightly lower fractions for stars 
in spheroids vs. disks (e.g. Schechter \& Dressler 1987;
Persic \& Salucci 1992), it is now generally accepted that the
majority of stars belong to spheroids. Figure 1 shows separately the mass
functions of color-selected ETGs and of blue, star-forming galaxies,
also based on the SDSS data (Baldry et al. 2004).  Above $\sim 3\times
10^{10}\msun$ red-sequence galaxies start to increasingly outnumber
blue galaxies by a factor that exceeds 10 above $\sim 3\times
10^{11}\msun$. Figure 2, drawn from the same mass functions, shows the
contributions to the total stellar mass by red and blue galaxies in
the various mass bins, along with the contributions to the total
number of galaxies. Then, ETGs represent only 17\% of the total number
of galaxies in the sample, but contribute $\sim 57\%$ of the total
mass. Moreover, $\gsim 80\%$ of the stellar mass in ETGs belongs to
galaxies more massive than $\sim 3\times 10^{10}\msun$. Dwarfs are
sometimes seen as the ``building blocks'' of galaxies, but at least in
the present universe one can not build much with them.

With spheroids holding the major share of stellar mass in
 galaxies, understanding their evolution -- from
 formation to their present state -- is central to the galaxy
 evolution problem in general. Historically, two main scenarios have
 confronted each other, the so-called Monolithic Collapse model
 (Eggen, Lynden-Bell \& Sandage 1962, Larson 1974, Arimoto \& Yoshii
 1987, Bressan, Chiosi \& Fagotto 1994), and the Hierarchical
 Merging model (e.g., Toomre 1977; White \& Rees 1978). In the
 former scenario spheroids form at a very early epoch as a result of a
 global starburst, and then passively evolve to the present.  If the
 local conditions are appropriate, a spheroid can gradually grow a
 disk by accreting gas from the environment, hence spheroids precede
 disks.  In the merging model, big spheroids result from the mutual
 disruption of disks in a merging event, hence disks precede
 spheroids.

The two scenarios appear to sharply contradict each other, but the
contradiction has progressively blurred in recent years. Evidence has
accumulated that the bulk of stars in spheroids are old, and most
likely formed in major merging events. In the hierarchical merging
scenario (the only one rooted in a solid cosmological context)
successive generations of models have struggled to increase their
predicted stellar ages, so as to produce results resembling the
opposite scenario.  This review will not attempt to trace an history
of theoretical efforts to understand the formation and evolution of
ellipticals and spheroids.  It will rather concentrate on reviewing
the accumulating observational evidences coming from the stellar
component of these galaxies. Other extremely interesting properties of
ellipticals, such as their structural and dynamical properties, their
hot gas content, central supermassive black holes, etc., will not be
touched in this review, even if they are certainly needed to complete
the picture and likely play an important role in the evolution of
these galaxies. Also untouched are the internal properties of ETGs,
such as color and line-strength gradients indicative of spatial
inhomogeneities of the stellar populations across the body of ETGs.  A
complementary view of ETG formation based on the globular cluster
populations of these galaxies is presented in the article by J. Brodie
\& J. Strader (this volume).

 The ``concordance cosmology'' ($\Omega_{\rm M}=0.3,\; \Omega_\Lambda
=0.7,\; H_\circ =70$) is adopted if not explicitly stated otherwise.
%\vfill\eject
\section{SYNTHETIC STELLAR POPULATIONS}

The ETGs treated in this review can only be studied in integrated light, 
hence the interpretation of their photometric and spectroscopic properties needs
population synthesis tools. Pioneering unconstrained synthesis
using ``quadratic programming'' (e.g., Faber 1972), was soon abandoned 
in favor of evolutionary population synthesis, whose foundations were 
laid down by Beatrice Tinsley in the 1970s (Tinsley \& Gunn 1976,
Tinsley 1980, Gunn, Stryker \& Tinsley 1981). Much progress has been made
in the course of the subsequent quarter of a century, especially thanks to the
systematic production of fairly complete libraries of stellar evolutionary
sequences and stellar spectra.

Several modern population synthesis tools are commonly in use today,
including those of Worthey (1994), Buzzoni (1995), Bressan, Chiosi, \&
Tantalo (1996), Maraston (1998, 2005), Bruzual \& Charlot (2003), Fioc
\& Rocca Volmerange (1997, PEGASE Code), V\'azques \& Leitherer (2005,
Starburst99 Code), Vazdekis et al. (2003), and Gonz\'alez Delgado et
al. (2005). Though far more reliable than earlier generations of 
models, even the most recent tools still may suffer from incomplete
spectral libraries (especially at high metallicity and for nonsolar
abundance ratios), and poorly calibrated mass loss in advanced stages,
such as the asymptotic giant branch (AGB). Yet, there is fair
agreement among the various models, with the exception of those for
ages around $\sim 1$ Gyr, when the contribution by AGB stars is at
maximum, and Maraston's models (calibrated on Magellanic Cloud
clusters) give appreciably higher near-IR fluxes than the other
models.

Only a few ``rules of thumb'' regarding population synthesis models
can be recalled here, which may be useful in guiding the reader
through some of the subtleties of their comparison with the
observations.

\pb No evolutionary population synthesis code is perfect. Evolutionary
tracks are not perfect and stellar libraries are never really
complete. So, any code deficiency will leave its imprint on the
results, generating a distortion of the age/metallicity grids used to
map plots of one observable versus another. Inevitably, such
distortions will leave their imprint in the results, and to some extent
may lead to spurious correlations/anticorrelations when reading ages
and metallicities from overplotted data points.

\pb 
Ages derived from best fits to simple
stellar populations (SSPs, i.e., single burst populations) are always
luminosity-weighted ages, and in general are more sensitive to the
youngest component of the real age distribution. SSP ages should be
regarded as lower limits.  

\pb Spectra and colors of SSPs are fairly
insensitive to the initial mass function (IMF), because most of the
light comes from stars in a narrow mass interval around the mass of
stars at the main sequence turnoff.  

\pb 
The time evolution of the
luminosity of a SSP does depend on the IMF, and so does the mass-to-light 
ratio ($M/L$). For example, a now fashionable IMF that flattens below $\sim
0.6\,\msun$ (e.g., Chabrier 2003) gives $M/L$ ratios a factor of $\sim
2$ lower than a straight Salpeter IMF. 

\pb Stellar ages and metallicities are the main quantities that the
analyses of colors and integrated spectra of galaxies are aimed to
determine. Yet, for many observables, age and metallicity are largely
degenerate, with a reduced age coupled to an increased metallicity
conjuring to leave the spectral energy distribution nearly
unchanged. This results primarily from the color (temperature) of the
main sequence turnoff, e.g., $(B-V)^{\rm TO}$, (the true clock of
SSPs) being almost equally sensitive to age and metallicity
changes. Indeed, from stellar isochrones one can derive that
$(\partial\log\, t/\partial[{\rm Fe/H}])_{(B-V)^{\rm TO}}\simeq -0.9
-0.35$[Fe/H], and a factor of 2 error in estimated metallicity
produces a factor $\sim 2$ error in age (Renzini 1992). Red giant
branch stars are the major contributors of bolometric luminosity in
old stellar populations, and their locus shifts to lower temperatures
with both increasing age and metallicity, further contributing to the
degeneracy.  Thus, from full SSPs, Worthey (1994) estimated that a
factor of 3 error in metallicity generates a factor of 2 error in age
when using optical colors as age indicators, the so-called 2/3
rule. Several strategies have been devised to circumvent this
difficulty and break the age-metallicity degeneracy (see below).

\pb
There are occasionally ambiguities in what is meant by the $M/L$ ratio
in the tabulated values. The mass $M$ can be defined either as the
mass of gas that went into stars, or the mass of the residual
population at age $t$, including the mass in dead remnants (i.e., the
original mass diminished by the mass lost by stars in the course of
their evolution), or even the mass of the surviving stars, i.e.,
without including the mass in remnants. Caution should be paid when
using tabular values, as different authors may adopt different
definitions.

The power of stellar population diagnostics stems from  the
opportunity to age-date the stellar content of galaxies in a fashion
that is independent of cosmological parameters. Then, once a cosmology
is adopted, ages derived from observations at a lower redshift can be
used to predict the properties of the stellar populations of ETGs at a
higher one, including their formation redshift. Thus, ages derived for
the local elliptical galaxies imply a well-defined color, spectral,
and luminosity evolution with redshifts, which all can be subject to
direct observational test. The extent to which a consistent picture of ETG
formation is emerging from low- and high-redshift observations is the
main underlying theme of this review.

\section{ELLIPTICAL GALAXIES IN THE LOCAL UNIVERSE}

Observations at high redshift are certainly the most direct way to
look at the forming galaxies, and a great observational effort is
currently being made in this direction. Yet, high-redshift galaxies
are very faint, and only few of their global properties can now be
measured. Nearby galaxies can instead be studied in far greater
detail, and their fossil evidence can provide a view of galaxy
formation and evolution that is fully complementary to that given by
high-redshift observations. By fossil evidence one refers to those
observables that are not related to ongoing, active star formation,
and which are instead the result of the integrated past star formation
history.  At first studies attempted to estimate ages and
metallicities of the dominant stellar populations on a
galaxy-by-galaxy basis.  But the tools used were still quite
rudimentary, being based on largely incomplete libraries of stellar
spectra and evolutionary sequences. Hence, through the 1980s
progress was relatively slow, and opinions could widely diverge as to
whether ellipticals were dominated by old stellar populations --as old
as galactic globular clusters-- or by intermediate age ones, several
billion years younger than globulars (see e.g. O'Connell 1986, Renzini 1986)
--with much of the diverging interpretations being a result of the
age-metallicity degeneracy. From the beginning of the 1990s
progress has been constantly accelerating, and much of this review
concentrates on the developments that took place over the past 15
years.

\subsection{Color-Magnitude Relation, Fundamental-Plane and Line-Indices}

3.1.1 THE COLOR-MAGNITUDE AND COLOR-$\sigma$ RELATIONS That
elliptical galaxies follow a tight color-magnitude (C-M) relation was
first recognized by Baum (1959), and in a massive exploration
Visvanathan \& Sandage (1977) and Sandage \& Visvanathan (1978a,b)
established the universality of this relation with what continues to be
the culmination of ETG studies in the pre-CCD era. The C-M relation
looked the same in all nine studied clusters, and much the same in the
field as well, though with larger dispersion (at least in part
due to larger distance errors). The focus was on the possible use of
the C-M relation as a distance indicator; however, Sandage \& Visvanathan
documented the tightness of the relation and noted that it
implies the stellar content of the galaxies to be very uniform. They also 
estimated that both S0's and ellipticals had to be evolving passively
since at least $\sim 1$ Gyr ago. Figure 3 shows a modern rendition for
the C-M plot for the Coma cluster galaxies, showing how tight it is,
as well as how closely both S0's and ellipticals follow the same relation,
as indeed Sandage \& Visvanathan had anticipated.

\smallskip\pn 

\begin{figure}[t]
%\vskip-0.05truecm
\centerline{\psfig{file=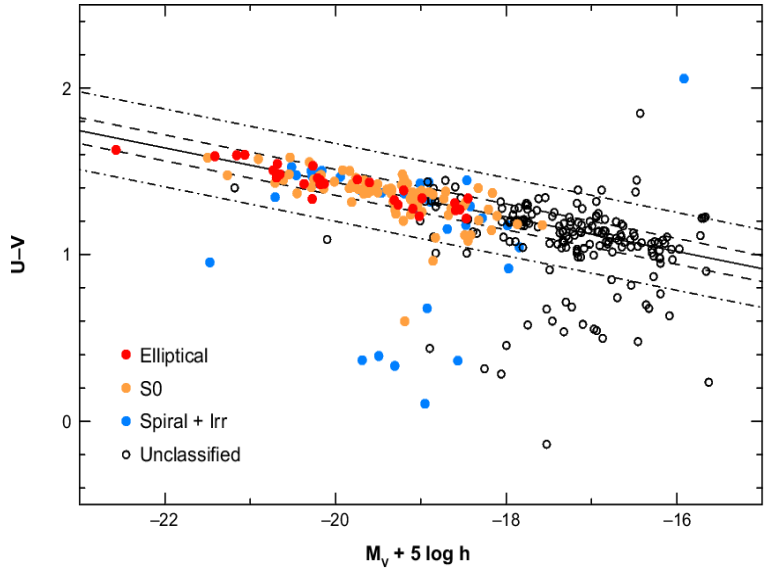,width=10cm,angle=-90,height=7.5cm}}

\vskip 1truecm
%\plotone{feab.eps}
\vskip-1truecm
\caption{\piedi The $(U-V)-M_{\rm V}$ color-magnitude relation for 
 galaxies of the various morphological types that are spectroscopic members of
the  Coma  cluster (Bower et al. 1999).
 }
\end{figure}

In a major breakthrough in galaxy dating, Bower, Lucey, \& Ellis
(1992), rather than trying to age-date galaxies one by one, were able
to set tight age constraints on all ETGs in Virgo and Coma at
once. Noting the remarkable homogeneity of ETGs in these clusters,
they estimated the intrinsic color scatter in the color-$\sigma$
relation (see Figure 4) to be $\delta(U-V)\lsim 0.04$ mag, where $\sigma$ is
the central stellar velocity dispersion of these galaxies. They
further argued that -- if due entirely to an age dispersion $\delta
t=\beta (t_{\rm H}-t_{\rm F}$), such color scatter should be equal to
the time scatter in formation epochs, times $\partial(U-V)/\partial
t$, i.e.:
\begin{equation}
t_{\rm H}-t_{\rm F}\;\lsim\; {0.04\over\beta}\Bigl({\partial (U-V)\over\partial
 t}\Bigr)^{-1},
\end{equation}
where $t_{\rm H}$ is the age of the universe at $z=0$, and galaxies
are assumed to form before a lookback time $t_{\rm F}$. Bower and colleagues
introduced the parameter $\beta$, such that $\beta(t_{\rm
H}-t_{\rm F})$ is the fraction of the available time during which
galaxies actually form. Thus, for $\beta=1$ galaxy formation is
uniformly distributed between $t\sim 0$ and $t=t_{\rm H}-t_{\rm F}$ ,
whereas for $\beta<1$ it is more and more synchronized, i.e., restricted
to the fraction $\beta$ of time interval $t_{\rm H}-t_{\rm F}$.
Adopting $\partial(U-V)/\partial t$ from the models of Bruzual (1983),
they derived $t_{\rm H}-t_{\rm F}< 2$ Gyr for $\beta =1$ and $t_{\rm H}-t_{\rm
F}< 8$ Gyr for $\beta =0.1$, corresponding respectively to
formation redshifts $\zf\gsim 2.8$ and $\gsim 1.1$ for their adopted
cosmology ($t_{\rm H}=15$ Gyr, $q_\circ=0.5$). For the concordance
cosmology, the same age constraints imply $\zf\gsim 3.3$ and $\gsim
0.8$, respectively. A value $\beta=0.1$ implies an extreme
synchronization, with all Virgo and Coma galaxies forming their stars
within less than 1 Gyr when the universe had half its present age,
which seems rather implausible. Bower and colleagues concluded that
ellipticals in clusters formed the bulk of their stars at $z\gsim 2$,
and later additions should not provide more than $\sim 10\%$ of
their present luminosity.  Making minimal use of stellar population
models, this approach provided for the first time a robust
demonstration that cluster ellipticals are made of very old stars,
with the bulk of them having formed at $z\gsim 2$.

\begin{figure}[t]
%\vskip-0.05truecm
\centerline{\psfig{file=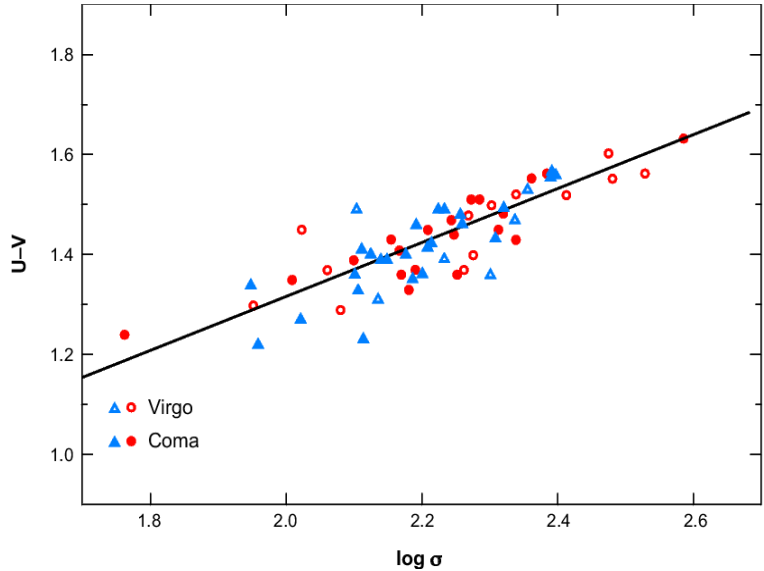,width=10cm,angle=-90,height=7.5cm}}

\vskip 1truecm
%\plotone{feab.eps}
\vskip-1truecm
\caption{\piedi The relation between the $(U-V)$ color and the central
velocity dispersion ($\sigma$) for early-type galaxies in the Virgo
(open symbols) and Coma (filled symbols) clusters. Red circles represent
ellipticals, blue triangles represent S0's. From Bower, Lucey \& Ellis (1992).
  }
\end{figure}

As the narrowness of the C-M and color-$\sigma$ relations sets
constraints on the ages of stellar populations in ETGs, their slope
can set useful constraints on the amount of merging that may have led
to the present-day galaxies. The reason is that merging without star
formation increases luminosity and $\sigma$, but leaves colors
unchanged, thus broadening and flattening the relations. Moreover,
merging with star formation makes bluer galaxies, thus broadening and
flattening the relations even more. Then, from the constraints set by
the slope of the C-M relation, Bower, Kodama, \& Terlevich (1998)
concluded that not only the bulk of stars in clusters must have formed
at high redshift, but also that they cannot have formed in mass units
much less than about half their present mass.

\smallskip\pn
3.1.2 THE FUNDAMENTAL PLANE Three key observables of
elliptical galaxies, namely the effective radius $\re$, the central
velocity dispersion $\sigma$, and the luminosity $L$ (or equivalently
the effective surface brightness $\ie=L/2\pi\re^2$) relate their
structural/dynamical status to their stellar content.  Indeed, elliptical
galaxies are not randomly distributed within the 3D space
($\re,\sigma,\ie$), but rather cluster close to a plane, thus known as
the fundamental plane (FP), with $\re\propto\sigma^{\rm a}\ie ^{\rm
b}$ (Dressler et al.  1987; Djorgovski \& Davis 1987), where the exponents
$a$ and $b$ depend on the specific band used for measuring the
luminosity. The projection of the FP over the ($\re,\ie$) coordinate
plane generates the Kormendy relation (Kormendy 1977), whereas a
projection over the ($\sigma, L=2\pi\re^2\ie$) plane generates the
Faber-Jackson relation (Faber \& Jackson 1976). At a time when testing
the $\Omega_{\rm M}=1$ standard cosmology had high priority, the FP
was first used to estimate distances, in order to map deviations
from the local Hubble flow and construct the gravitational potential
on large scales. Its use to infer the properties of the stellar
content of galaxies, and set constraints on their formation, came
later.  Yet, by relating the luminosity to the structural-dynamical
parameters of a galaxy, the FP offers a precious tool to gather
information on the ages and metallicities of galaxies, at low as well
as at high redshifts.

The mere existence of a FP implies that ellipticals ({\it a}) are
virialised systems, ({\it b}) have self-similar (homologous) structures, or 
their structures
(e.g., the shape of the mass distribution) vary in a systematic
fashion along the plane, and ({\it c}) contain stellar populations which
must fulfill tight age and metallicity constraints.  Here we
concentrate on this latter aspect. 
%(It is not surprizing that
%ellipticals are well virialized, given that their dynamical time is
%much shorter than their age.)

To better appreciate the physical implications of the FP, Bender,
Burnstein, \& Faber (1992) introduced an orthogonal coordinate system
($\ku ,\kd ,\kt$), in which each new variable is a linear combination
of $\log\sigma^2,\,\log\re$ and $\log\ie$. The transformation
corresponds to a rotation of the coordinate system such that in the
$(\ku ,\kt$) projection the FP is seen almost perfectly
edge-on. Moreover, if structural homology holds all along the plane,
then $\log\,M/L=\sqrt3\kt$ + const. If $\sigma$ is (almost) unaffected by the
dark matter distribution (as currently understood, Rix et al. 1997),
then $\kt$ provides a measure of the stellar $M/L$ ratio, and
$\ku\propto\log(\sigma^2\re)\propto \log\, M$ a measure of the stellar
mass. Bender and colleagues showed that in Virgo and Coma the FP is
remarkably ``thin'', with a 1-$\sigma$ dispersion perpendicular to the
plane of only $\sigma(\kt)\simeq 0.05$, corresponding to a dispersion
in the $M/L$ ratio $\lsim 10\%$ at any position along the
plane. Moreover, the FP itself is ``tilted'', with the $M/L$ ratio
apparently increasing by a factor $\sim 3$ along the plane, while the
mass is increasing by a factor $\sim 100$. Note that the tilt does not
imply a departure from virialization, but rather a systematic trend of
the stellar content with galaxy mass, possibly coupled with a
systematic departure from structural homology (e.g., Bender, Burstein
\& Faber 1992, Ciotti 1997, Busarello et al. 1997).

The narrowness of the FP, coupled to the relatively large tilt
($\Delta\kt/\sigma(\kt)\simeq 0.35/0.05=7$) requires some sort of fine
tuning, which is perhaps the most intriguing property of
the FP (Renzini \& Ciotti 1993).  Although unable to identify one
specific origin for the FP tilt, Renzini \& Ciotti argued that
the small scatter perpendicular to the FP implied a small age
dispersion ($\lsim 15\%$) and high formation redshift, fully
consistent with the Bower, Lucey \& Ellis (1992) argument based on the
narrowness of the C-M and color-$\sigma$ relations.

The remarkable properties of the FP for the Virgo and Coma clusters
were soon shown to be shared by all studied clusters in the local
universe. J{\o}rgensen, Franx, \& Kj{\ae}rgaard (1996) constructed the
FP for 230 ETGs in 10 clusters (including Coma), showing that the FP
tilt and scatter are just about the same in all local clusters, thus
strengthening the case for the high formation redshift of cluster ETGs
being universal. However, Worthey, Trager, \& Faber (1995) countered
that the thinness of the FP, C-M, and color-$\sigma$ relations could
be preserved, even with a large age spread, provided age and
metallicity are anticorrelated (with old galaxies being metal poor
and young ones being metal rich). This is indeed what Worthey and colleagues
reported from their line-indices analysis (see below), indicating
a factor of $\sim 6$ for the range in age balanced by a factor $\sim 10$ in
metallicity (from solar to $\sim 10$ times solar). If so, then the FP
should be thicker in the near infrared, because the compensating
effect of metallicity would be much lower at longer wavelength, thus
unmasking the full effect of a large age spread (Pahre, Djorgovski, \&
De Carvalho 1995). But Pahre and colleagues found the scatter of the FP
$K$-band to be the same as in the optical. In addition, its slope
implied a sizable variation of $M/\lk\propto M^{0.16}$ along the FP,
somewhat flatter than in the optical ($M/\lv\propto M^{0.23}$), still
far from the $M/\lk\sim$const. predicted by Worthey et al. (1995).

These conclusions were further documented and reinforced by Pahre,
Djorgovski, \& De Carvalho (1998), Scodeggio et al. (1998), Mobasher
et al. (1999), and Pahre, De Carvalho, \& Djorgovski (1998), who
finally concluded that the origin of the FP tilt defies a simple
explanation, but is likely the result of combined age and metallicity
trends along the plane (with the most metal rich galaxies being
actually the oldest), plus an unidentified systematic deviation from
structural homology. Several possibilities for the homology breaking
have been proposed and investigated, such as variation in stellar
and/or dark matter content and/or distribution, anisotropy, and
rotational support (e.g., Ciotti, Lanzoni, \& Renzini 1996, Prugniel
\& Simien 1996, Ciotti \& Lanzoni 1997). Recently, Trujillo, Burkert,
\& Bell (2004) argued that one fourth of the tilt is due to stellar
population (i.e., a combination of metallicity and age), and three
quarters of it to structural nonhomology in the distribution of the
visible matter.

Of special interest is the comparison of the FP in clusters and in the
field, because one expects all formation processes to be faster in
high density peaks of the matter distribution. This was tested by
Bernardi et al. (2003b, 2006) with a sample of $\sim 40,000$ SDSS
morphology- and color-selected ETGs spanning a wide range of
environmental conditions, from dense cluster cores to very low
densities. Bernardi and colleagues found very small, but detectable
differences in the FP zero point; the average surface brightness is
$\sim 0.08$ mag brighter at the lowest density extreme compared to the
opposite extreme. As the sample galaxies are distributed in redshift
up to $z\sim 0.3$, they used the observed lookback time to empirically
determine the time derivative of the surface brightness (hence in a
model-independent fashion) and estimated that the 0.08 mag difference
in surface brightness implies an age difference of $\sim 1$ Gyr, and
therefore that galaxies in low density environments are $\sim 1$ Gyr
younger compared to those in cluster cores.
\smallskip\pn
3.1.3 THE LINE-STRENGTH DIAGNOSTICS Optical spectra of
ETGs present a number of absorption features whose strength must
depend on the distributions of stellar ages, metallicities and
abundance ratios, and therefore may give insight over such
distributions.  To exploit this opportunity, Burstein et al. (1984)
introduced a set of indices now known as the Lick/IDS system, and started
taking measurements for a number of galaxies. The most widely used indices
have been the $\mgt$ (or $\mgb$), $\fe$, and the $\hbe$ indices,
measuring respectively the strength of MgH+MgI at $\lambda\simeq
5156-5197$\AA, the average of two FeI lines at $\lambda\simeq 5248$
and $5315$\AA, and of H$_\beta$.

A first important result was the discovery that theoretical
models based on solar abundance ratios adequately describe the
combinations of the values of the $\fe$ and $\mgt$ indices in
low-luminosity ETGs, but fail for bright galaxies (Peletier 1989,
Gorgas, Efstathiou, \& Arag{\'o}n-Salamanca 1990, Faber, Worthey, \&
Gonzales 1992, Worthey, Faber, \& Gonzales 1992, Davies, Sadler, \&
Peletier 1993, J{\o}rgensen 1997).  This implies either that
population synthesis models suffered from some inadequacy at high
metallicity (possibly due to incomplete stellar libraries), or that
massive ellipticals were genuinely enriched in magnesium relative to
iron, not unlike the halo stars of the Milky Way  (e.g., Wheeler,
Sneden, \& Truran 1989). As for the Milky Way  halo, such an $\alpha$-element
overabundance may signal a prompt enrichment in heavy elements from
Type II supernovae, with the short star-formation timescale having
prevented most Type Ia supernovae from contributing their iron while
star formation was still active. Yet, a star-formation timescale
decreasing with increasing mass was contrary to the expectations of
galactic wind/monolithic models (e.g., Arimoto \& Yoshii 1987), where
the star formation timescale increases with the depth of the potential
well (Faber, Worthey \& Gonzales 1992). However, as noted by Thomas
(1999), the contemporary semi-analytical models did not predict any
$\alpha$-element enhancement at all, no matter whether in low- or
high-mass ETGs.  Indeed, Thomas, Greggio, \& Bender (1999) argued that
the $\alpha$-enhancement, if real, was also at variance with a
scenario in which massive ellipticals form by merging spirals, and
required instead that star formation was completed in less than $\sim
1$ Gyr. Therefore, assessing whether the $\alpha$-enhancement was
real, and in that case measuring it, had potentially far reaching
implications for the formation of ETGs.

Two limitations had to be overcome in order to reach a credible
interpretation of the $\fe -\mgb$ plots: ({\it a}) existing synthetic
models for the Lick/IDS indices were based on stellar libraries with
fixed [$\alpha$/Fe] (Worthey 1994, Buzzoni 1995), and ({\it b}) an
empirical verification of the reality of the $\alpha$-enhancement was
lacking. In an attempt to overcome the first limitation, Greggio
(1997) developed a scaling algorithm that allowed one to use existing
models with solar abundance ratios to estimate the Mg overabundance,
and she concluded that an enhancement up to [Mg/Fe]$\simeq +0.4$ was
required for the nuclei of the most massive ellipticals (see also
Weiss, Peletier, \& Matteucci 1995). She also concluded that a
closed-box model for chemical evolution failed to explain the very
high values of the $\mgt$ index of these galaxies. Indeed, the
numerous metal-poor stars predicted by the model would obliterate the
$\mgt$ feature, hence the nuclei of ellipticals had to lack
substantial numbers of stars more metal poor than $\sim
0.5\zsun$. Besides, very old ages ($\gsim 10$ Gyr) and
$\alpha$-enhancement were jointly required to account for galaxies
with strong $\mgt$.  Eventually, Thomas, Maraston, \& Bender (2003)
produced a full set of synthetic models with variable [$\alpha$/Fe],
and Maraston et al. (2003) compared such models to the indices of ETGs
and of metal-rich globular clusters of the Galactic bulge, for which
the $\alpha$-enhancement has been demonstarted on a star-by-star basis
by high resolution spectroscopy. The result is displayed in Figure 5,
showing that indeed the new models indicate for the bulge globulars an
enhancement of [$\alpha$/Fe]$\sim +0.3$, in agreement with the stellar
spectroscopy results, and similar to that indicated for massive ETGs.

\begin{figure}[t]
\vskip-0.05truecm
\centerline{\psfig{file=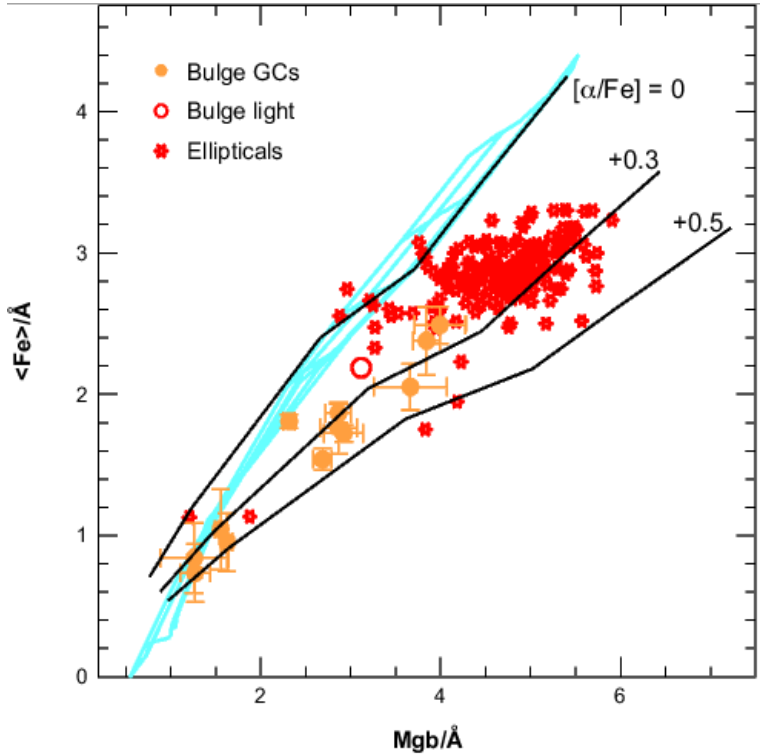,width=10cm,angle=0,height=10cm}}

\vskip 1truecm
%\plotone{feab.eps}
\vskip-1truecm
\caption{\piedi The $\fe$ index versus the $\mgb$ index for a sample
of halo and bulge globular clusters, the bulge integrated light in
Baade's Window, and for elliptical galaxies from various
sources. Overimposed are synthetic model indices (from Thomas,
Maraston \& Bender 2003) with solar metallicity ([Z/H]=0), various
$\alpha$-enhancements as
indicated, and an age of 12 Gyr (black solid lines).  The cyan grid
shows a set of simple stellar population models (from Maraston 1998)
with solar abundance ratios, metallicities from [Fe/H] = --2.25 to
+0.67 (bottom to top), and ages from 3 to 15 Gyr (left to right).
The blue grid offers an example of the so-called age-metallicity
degeneracy. From Maraston et al. (2003). 
}
\end{figure}

Other widely used diagnostic diagrams involved the $\hbe$ index along
with $\fe$ and $\mgt$ or $\mgb$. The Balmer lines had been suggested
as good age indicators (e.g., O'Connell 1980; Dressler \& Gunn 1983),
an expectation that was confirmed by the set of synthetic models
constructed by Worthey (1994) with the aim of breaking the
age-metallicity degeneracy that affects the broad-band colors of
galaxies. Worthey's models were applied by J{\o}rgensen (1999) to a
sample of 115 ETGs in the Coma cluster, and by Trager et al. (2000) to
a sample of 40 ETGs biased toward low-density environments, augmented
by 22 ETGs in the Fornax cluster from Kuntschner \& Davies (1998),
which showed systematically lower $\hbe$ indices. From these samples,
and using the $\hbe-\mgb$ and $\hbe-\fe$ plots from Worthey's models,
both J{\o}rgensen and Trager and colleagues concluded that ages ranged
from a few to almost 20 Gyr, but age and metallicity were
anticorrelated in such a way that the $\mgb -\sigma$, C-M, and FP
relations may be kept very tight. Moreover, there was a tendency for
ETGs in the field to appear younger than those in clusters.  Yet,
Trager and colleagues cautioned that $\hbe$ is most sensitive to even
low levels of recent star formation, and suggested that the bulk of
stars in ETGs may well be old, but a small ``frosting'' of younger
stars drives some galaxies toward areas in the $\hbe-\mgb$ and
$\hbe-\fe$ plots with younger SSP ages. Finally, for the origin of the
$\alpha$-enhancement Trager and colleagues favored a tight correlation
of the IMF with $\sigma$, in the sense of more massive galaxies having
a flatter IMF, hence more Type II supernovae. However, with a flatter
IMF more massive galaxies would evolve faster in luminosity with
increasing redshift, compared to less massive galaxies, which appears
to be at variance with the observations (see below).

These conclusions had the merit of promoting further debates. Maraston
\& Thomas (2000) argued that even a small old, metal-poor component
with a blue horizontal branch (like in galactic globulars) would
increase the $\hbe$ index thus making galaxies look significantly
younger than they are. Even more embarrassing for the use of the
$\hbe-\mgb$ and $\hbe-\fe$ plots is that a perverse circulation of the
errors automatically generates an apparent anticorrelation of age and
metallicity, even where it does not exist. For example, if $\hbe$ is
overestimated by observational errors, then age is underestimated,
which in turn would reduce $\mgb$ below the observed value unless the
younger age is balanced by an artificial increase of metallicity.
Trager and colleagues were fully aware of the problem, and concluded
that only data with very small errors could safely be used. Kuntschner
et al. (2001) investigated the effect by means of Monte Carlo
simulations, and indeed showed that much of the  apparent
age-metallicity anticorrelation is a mere result of the tight
correlation of their errors.  They concluded that only a few outliers
among the 72 ETGs in their study are likely to have
few-billion-year-old luminosity-weighted ages, and these were
typically galaxies in the field or loose groups, whereas a uniformly
old age was derived for the vast majority of the studied
galaxies. Moreover, younger ages were more frequently indicated
for S0 galaxies (Kuntschner \& Davies 1998). Nevertheless, in a
cluster with very tight C-M and FP relations such as Coma, a large age
spread at all magnitudes was found for a sample of 247 cluster members
(Poggianti et al. 2001), and a sizable age-metallicity anticorrelation
was also found for a large sample of SDSS galaxies (Bernardi et
al. 2005).

As already alluded to, the main pitfall of the procedure is that the
various indices depend on all three population parameters one is
seeking to estimate: thus $\hbe$ is primarily sensitive to age, but
also to [Fe/H] and [$\alpha$/Fe], $\fe$ is sensitive to [Fe/H], but
also to age and [Mg/Fe]; etc.  Thus, the resulting errors in age,
[Fe/H] and [Mg/Fe] are all tightly correlated, and one is left with
the suspicion that apparent correlations or anticorrelations may be an
artifact of the procedure, rather than reflecting the real properties
of galaxies.  In an effort to circumvent these difficulties, Thomas et
al. (2005) renounced to trust the results galaxy by galaxy. They
rather looked at patterns in the various index-index plots and
compared them to mock galaxy samples generated via Monte Carlo
simulations that fully incorporated the circulation of the errors.
The real result was not a set of ages and metallicities assigned to
individual galaxies, but rather age and metallicity trends with
velocity dispersion, mass and environments.  Having analyzed a sample
of 124 ETGs in high- and low-density environments, Thomas and colleagues
reached the following conclusions: ({\it a}) all three parameters --age,
metallicity and [$\alpha$/Fe]-- correlate strongly with $\sigma$, and, on
average, follow the relations:
\begin{equation}
\log t/{\rm Gyr} = 0.46\, (0.17) + 0.238\, (0.32)\,\log\,\sigma ,
\end{equation}
%$\;$ 
%\vskip -1.2 truemm
%\vskip -2.4 truemm
\begin{equation}
{\rm [Z/H]}=-1.06\, (-1.03) + 0.55\, (0.57)\, \log\,\sigma ,
\end{equation}
\begin{equation}
[\alpha/{\rm Fe}]=-0.42\, (-0.42) + 0.28\,(0.28)\,\log\,\sigma ,
\end{equation}
where quantities in brackets/not in brackets refer to
low-density/high-density environments, respectively.  ({\it b}) For
ETGs less massive than $\sim 10^{10}\msun$ there is evidence for the
presence of intermediate-age stellar populations with near-solar
Mg/Fe.  Instead, massive galaxies ($\gsim 10^{11}\msun$) appear
dominated by old stellar populations, whereas at intermediate masses
the strength of $\hbe$ requires either some intermediate age component
or a blue horizontal branch (HB) contribution.  ({\it c}) By and
large this picture applies to both cluster and field ETGs, with
cluster galaxies having experienced the bulk of their star formation
between $z\sim 5$ and 2, and this activity appears to have been
delayed by $\sim 2$ Gyr in the lowest density environments, i.e.,
between $z\sim 2$ and $\sim 1$. Figure 6 qualitatively summarizes this
scenario, in which the duration of star formation activity decreases
with increasing mass (as required by the [Mg/Fe] trend with $\sigma$),
and extends to younger ages for decreasing mass (as forced by the
$\hbe -\sigma$ relation).  Note that the smooth star-formation
histories in this figure should be regarded as probability
distributions, rather than as the actual history of individual
galaxies, where star formation may indeed take place in a series of
bursts. Qualitatively similar conclusions were reached by Nelan et
al. (2005), from a study of $\sim 4000$ red-sequence galaxies in $\sim
90$ clusters as part of the National Astronomical Observatory
Fundamental Plane Survey. Assuming the most massive galaxies
($\sigma\sim 400\,\kms$) to be 13 Gyr old, they derived an age of only
5.5 Gyr for less massive galaxies ($\sigma\sim 100\,\kms$). Note that
the age-$\sigma$ scaling of Thomas and colleagues would have given a much
older age ($\sim 9.5$ Gyr). Taken together, Equations 2 and 3 imply a trend
of $M/\lv$ by a factor $\sim 1.8$ along the FP (from $\sigma=100$ to
350 \kms), thus accounting for almost two thirds of the FP tilt.

As extensively discussed by Thomas et al. (2005), one residual concern 
comes from the possibility that part of the $\hbe$ strength 
may be  due to blue HB  stars. Besides a blue HB contribution
by low-metallicity stars (especially in less massive galaxies), blue HB stars
may also be produced by old, metal-rich populations, and appear to be 
responsible for the UV upturn in the spectrum of local ETGs (Brown et al. 2000,
Greggio \& Renzini 1990). In the Thomas et al. sample, some S0 outliers
with strong $\hbe$ and strong metal lines would require very young ages and 
extremely high metallicity (up to $\sim 10$ times solar), and may better be 
accounted for by an old, metal-rich  population with a well-developed blue HB.

\begin{figure}[t]
%\vskip-0.55truecm
\centerline{\psfig{file=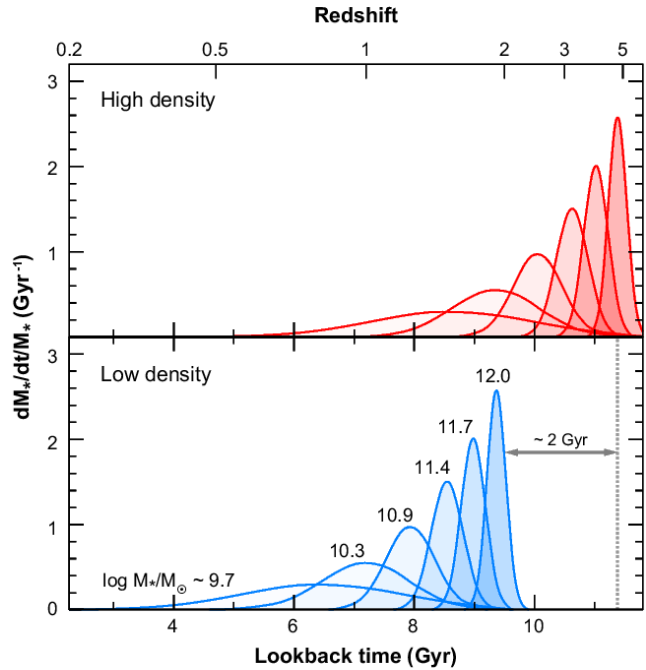,width=10cm,angle=-90,height=9cm}}

\vskip 1truecm
%\plotone{feab.eps}
\vskip-1truecm
\caption{\piedi The scenario proposed by Thomas et al. (2005) for the
average star formation history of early-type galaxies of different
masses, from $5\times 10^9\msun$ up to $10^{12}\msun$, corresponding
to $\sigma\simeq 100$ to $\sim 320\,\kms$, for the highest and lowest
environmental densities, respectively, in the upper and lower panel.  
}
\end{figure}

The $\mgt-\sigma$ relation has also been used to quantify
environmental differences in the stellar population content. The
cluster/field difference turns out to be small, with $\Delta\mgt\sim
0.007$ mag, corresponding to $\sim 1$ Gyr difference -- field galaxies
being younger -- within a sample including $\sim 900$ ETGs (Bernardi
et al. 1998), though no statistically significant environmental
dependence of both $\mgt$ and $\hbe$ was detected within a sample of
$\sim 9,000$ ETGs from the SDSS (Bernardi et al. 2003a). Still from
SDSS, coadding thousands of ETG spectra in various luminosity and
environment bins, Eisenstein et al. (2003) detect clear trends with
the environment thanks to the resulting exquisite S/N, but the
differences are very small, and Eisenstein and colleagues refrain from
interpreting them in terms of age/metallicity differences.

These results from the analysis of the Lick/IDS indices, including
large trends of age with $\sigma$, or even large age-metallicity
anticorrelations, have yet to be proven consistent with the FP and C-M
relations of the same galaxies as established specifically for the
studied clusters. Feeding the values of the indices, the synthetic
models return ages, metallicities, and $\alpha$-enhancements.  But
along with them the same models also give the colors and the stellar
$M/L$ ratio of each galaxy in the various bands, hence allowing one to
construct implied FP and C-M relations. It would be reassuring for
the soundness of the whole procedure if such relations were found to
be consistent with the observed ones.  To our knowledge, this sanity
check has not been attempted yet.  The mentioned trends and
correlations, if real, would also have profound implications for the
evolution of the FP and C-M relations with redshift, an opportunity
that will be exploited below.

%\vfill \pn
%\smallskip\pn
%\vfill
%D. Radial Gradients

\subsection{Ellipticals Versus Spiral Bulges}
\begin{figure}[t]
%\vskip-0.5truecm
\centerline{\psfig{file=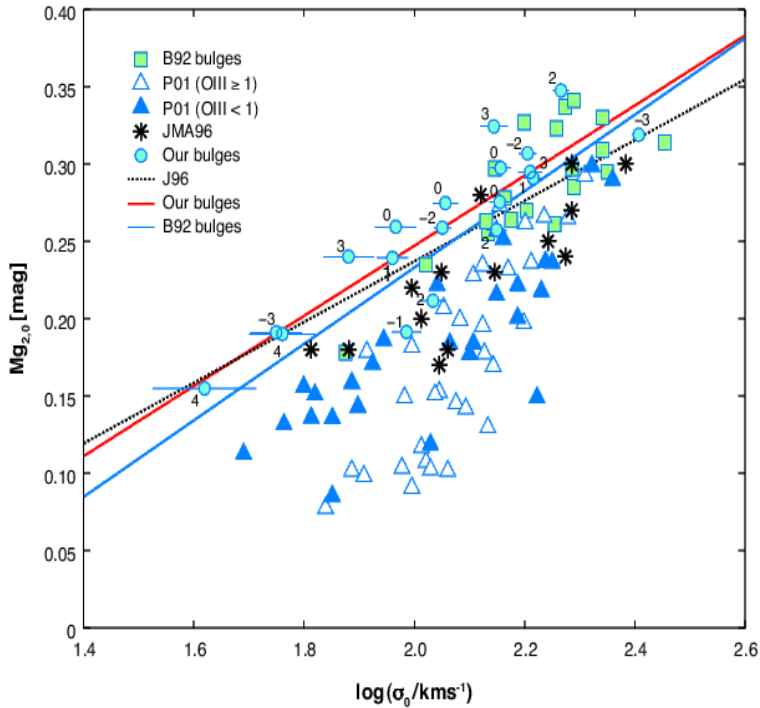,width=10cm,angle=-90,height=9.5cm}}

%\vskip-5truecm
%\plotone{feab.eps}
%\vskip 1truecm
\caption{\piedi The \mgs\ relation for the spiral bulges studied by
Falcon-Barroso et al. (2002), labelled ``our bulges'' in the insert,
are compared to bulges in other samples (B92 = Bender, Burnstein \&
Faber 1992, P01 = Prugniel, Maubon \& Simien 2001, JMA96 = Jablonka,
Martin \& Arimoto 1996). The solid lines are the bet fits to the
corresponding data, while the dashed line show the average relation
for cluster ellipticals from J{\o}rgensen, Franx \& Kj{\ae}rgaard  (1996).  
}
\end{figure}
The bulges of spiral galaxies are distinguished in ``true bulges'',
typically hosted by S0-Sb galaxies, and ``pseudobulges'' usually (but
not exclusively) in later-type galaxies (Kormendy \& Kennicutt
2004). True (classical) bulges have long been known as similar to
ellipticals of comparable luminosity, in both structure, line
strengths and colors (e.g., Bender, Burnstein \& Faber 1992, Jablonka,
Martin, \& Arimoto 1996, Renzini 1999, and references therein).
Peletier et al. (1999) were able to quantify this similarity using {\it Hubble
Space Telescope} (HST) WFPC2 (Wide Field Planetary Camera 2) and
NICMOS (Near Infrared Camera and Multiobject Spectrometer)
observations, and concluded that most (true) bulges in their sample of
20 spirals (including only 3 galaxies later than Sb) had optical and
optical-IR colors similar to those of Coma ellipticals. Hence, like in
Coma ellipticals their stellar populations formed at $z\gsim 3$, even
if most of the galaxies in their sample are in small groups or in the
field. More recently, Falcon-Barroso, Peletier, \& Balcells (2002)
measured the central velocity dispersion for the same sample observed
by Peletier and colleagues, and constructed the FP for these
bulges, showing that bulges in
this sample tightly follow the same FP relation as cluster
ellipticals, and therefore had to form their stars at nearly the same
epoch.  The similarity of true bulges and ellipticals includes the
tendency of less massive objects to have experienced recent star
formation, as indicated by their location in the $\mgt-\sigma$ diagram
in Figure 7. These
similarities between true (classical) bulges and ellipticals suggest a
similar origin, possibly in merger-driven starbursts at high
redshifts. Pseudobulges, instead, are more likely to have originated
via secular evolution of disks driven by bars and other deviations
from axial symmetry, as extensively discussed and documented by
Kormendy \& Kennicutt (2004). Several of the objects in the Prugniel
and colleagues sample in Figure 7 are likely to belong to the
pseudobulge group. From the Lick/IDS indices of a sample of bulges,
Thomas and Davies (2006) argue that the same scenario depicted in
Figure 6 for ETGs, applies to bulges as well, the main difference
being that bulges are on average less massive, hence on average
younger than ETGs.

Looking near to us, HST and ground based photometry of individual stars
in the Galactic bulge have shown that they are older than at least 10
Gyr, with no detectable trace of an intermediate age component
(Ortolani et al. 1995, Kuijken \& Rich 2002, Zoccali et
al. 2003). HST/NICMOS photometry of stars in the bulge of M31 has also
shown that their $H$-band luminosity function is virtually identical
to that of the Galactic bulge, and by inference should have
nearly identical ages (Stephens et al. 2003). These two bulges belong
to spirals in a rather small group, and yet appear to have formed
their stars at an epoch corresponding to $z\gsim 2$, not unlike most
ellipticals.

\subsection{Summary of the Low-Redshift (Fossil) Evidence}

The main observational constraints on the epoch of formation of the
stellar populations of ETGs in the near universe can be summarized as
follows: 
\pb The C-M, color-$\sigma$ and FP relations for ETGs in
clusters indicate that the bulk of stars in these galaxies formed at
$z\gsim 2-3$.  
\pb The same relations for the field ETGs suggest that
star formation in low density environments was delayed by $\sim 1-2$
Gyr.  
\pb The more massive galaxies appear to be enhanced in Mg
relative to iron, which indicates that the duration of the
star-formation phase decreases with increasing galaxy mass, having
been shorter than $\sim 1$ Gyr in the most massive galaxies.  
\pb
Interpretations of the Lick/IDS indices remains partly controversial,
with either an age-metallicity anticorrelation, or an increase of
both age and metallicity with increasing $\sigma$.

These trends are qualitatively illustrated in Figure 6, showing that the
higher the final mass of the system, the sooner star formation starts
and more promptly subsides, in an apparently ``antihierarchical''
fashion. A trend in which the stellar population age and metallicity
are tightly correlated to the depth of the potential well (as measured
by $\sigma$) argues for star formation, metal enrichment, supernova
feedback, merging, and violent relaxation having been all concomitant
processes rather than having taken place sequentially.  

The fossil evidence illustrated so far is in qualitative agreement
with complementary evidence at low as well as high redshift, now
relative to star-forming galaxies as opposed to quiescent ones. At low
$z$, Gavazzi (1993) and Gavazzi, Pierini, \& Boselli (1996) showed
that in local (disk) galaxies the specific star-formation rate
anticorrelates with galaxy mass, a trend that can well be extended
to include fully quiescent ellipticals. On this basis, Gavazzi and
collaborators emphasized that mass is the primary parameter
controlling the star-formation history of galaxies, with a sharp
transition at $L_{\rm H}\simeq 2\times 10^{10}\lsun$ (corresponding to
$\sim 2\times 10^{10}\msun$) between late-type, star-forming galaxies and
mostly passive, early-type galaxies (Scodeggio et al. 2002). This
transition mass has then been precisely located at $\sim 3\times
10^{10}\msun$ with the thorough analysis of the SDSS database
(Kauffmann et al. 2003). In parallel, high redshift observations have
shown that the near-IR luminosity (i.e., mass) of galaxies undergoing
rapid star formation has declined monotonically from $z\sim 1$ to the
present, a trend for which Cowie et al. (1996) coined the term
down-sizing.  This is becoming a new paradigm for galaxy
formation, as the anticorrelation of the specific star-formation rate
with mass is now recognized to persist well beyond $z\sim 2$ (e.g.,
Juneau et al. 2005, Feulner et al. 2006).

\section{ELLIPTICAL GALAXIES AT HIGH REDSHIFT}

Perhaps the best way of breaking the age-metallicity degeneracy is by
looking back in time, studying galaxies at higher and higher
redshifts. In the 1990s this was attempted first with 4m-class
telescopes, and later, with impressive success, with 8-10m-class
telescopes and HST.  Studies first focused on cluster ellipticals, and
their extension to field galaxies followed with some
delay. Thus, the evolution with redshift of various galaxy properties
were thoroughly investigated, such as the C-M and Kormendy relations,
the luminosity and mass functions, and the FP.  Various
ongoing surveys are designed to map the evolution with redshift
and local environment of all these properties, along with the number
density of these galaxies.

\subsection{Cluster Ellipticals Up to $z\sim 1$}

4.1.1 THE COLOR-MAGNITUDE RELATION
 With the
identification of clusters at higher and higher redshifts, from the
mid-1990s it became possible to construct their C-M relation,
hence to directly assess the rate of evolution of cluster
ETGs. Pioneering studies showed a clearly recognizable red sequence 
in high-redshift clusters, and gave hints that the color evolution
up to $z\sim 1$ was broadly consistent with pure
passive evolution of the galaxies formed at high redshift
(Dressler \& Gunn 1990; Arag{\'o}n-Salamanca et al. 1993; Rakos \&
Schombert 1995). Subsequent studies fully confirmed these early hints
and provided accurate estimates for the formation redshift of the bulk
of stars in cluster ellipticals. Thus, replicating the Bower, Lucey \& Ellis
(1992) procedure for a sample of morphologically-selected
ETGs in clusters at $z\sim 0.5$, Ellis et al. (1997)
were able to conclude that most of the star formation in ellipticals
in dense clusters was completed 5--6 Gyr earlier than the cosmic time
at which they are observed, i.e., at $z\gsim 3$.  Extending these
studies to clusters up to $z\sim 0.9$, Stanford, Eisenhardt, and Dickinson
(1998) showed that pure passive evolution continues all the
way to such higher redshift, while the dispersion of the C-M relation
remains as small as it is in Virgo and Coma (see Figure 8). Thus,
Stanford and colleagues concluded that cluster ellipticals formed the bulk of
their stars at $z\gsim 3$, with the small color dispersion arguing for
highly synchronized star-formation histories among galaxies within
each cluster, and from one cluster to another. These conclusions were
reinforced by several other investigators, e.g., Gladders et
al. (1998), Kodama et al. (1998), Nelson et al. (2001), De Lucia et
al. (2004), and by van Dokkum et al. (2000), who also cautioned
about the ``progenitor bias'' (see below).

\begin{figure}[t]
%\vskip-3.5truecm
\centerline{\psfig{file=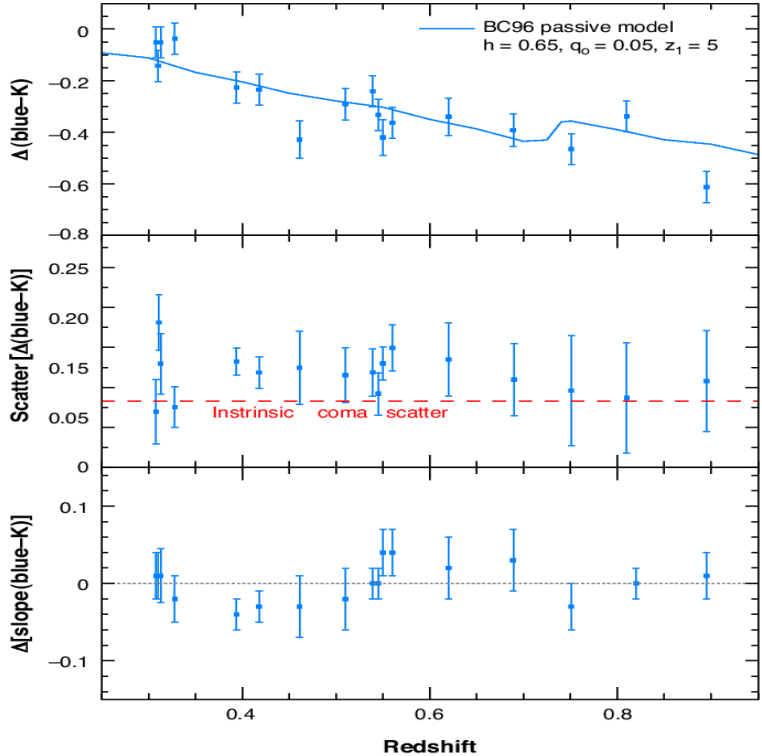,width=10cm,angle=0,height=10cm}}

%\vskip 1truecm
%\plotone{feab.eps}
\vskip 0.5truecm
\caption{\piedi The color evolution of early-type galaxies in clusters
out to $z\simeq 0.9$ (Stanford, Eisenhardt \& Dickinson 1998,
Dickinson 1997).  The ``blue'' band is tuned for each cluster to
approximately sample the rest frame $U$-band, whereas the $K$ band is
always in the observed frame. Top panel: the redshift evolution of the
zero point of ``blue''$-K$ color-magnitude (C-M) relation relative to
the Coma cluster.  A purely passive evolution model is also shown,
assuming a formation redshift $z_{\rm F}=5$.  Middle panel: the
intrinsic color scatter, having removed the mean slope of the
C-M relation in each cluster and the contribution of
photometric errors. The intrinsic scatter of Coma galaxies is shown
for reference. Bottom panel: the redshift evolution of the slope of
the (``blue''$-K)-K$ C-M relation, relative to  the slope of the corresponding 
relation for galaxies in Coma.  }
\end{figure}

The evolution of the C-M relation was then traced beyond $z=1$ thanks
to the discovery of higher redshift clusters, primarily by the Rosat
Deep Cluster Survey (RDCS, Rosati et al. 1998). Using deep HST/ACS
(Advanced Camera for Surveys)
$i$- and $z$-band images, Blakeslee et al. (2003) found a tight red
sequence for morphologically selected ETGs in a $z=1.24$ cluster,
implying typical ages of $\sim 3$ Gyr, and formation redshift
$\zf\gsim 3$. This was further confirmed by its infrared C-M relation
(Lidman et al. 2004).  However, some clusters in the range
$0.78<z<1.27$ appear to have a larger color scatter than others, 
again with ellipticals in those with tight C-M relation having virtually
completed their star formation at $z\gsim 3$ (Holden et
al. 2004). Presently, the highest redshift clusters known to be
dominated by old, massive ETGs are at $z\simeq 1.4$
(Stanford et al. 2005, Mullis et al. 2005).
 
The redshift evolution of the color of the red sequence in clusters
proved to be a very powerful tool in disentangling ambiguities that
are difficult to eliminate based only on $z\sim 0$ observations.  From
the color evolution of the red sequence in two Abell clusters at
$z\sim 0.2$ and $\sim 0.4$, Kodama \& Arimoto (1997) were able to
break the age-metallicity degeneracy plaguing most of the global
observables of local ellipticals. In principle, because colors depend
both on age and metallicity, the slope of the C-M relation could
equally well be reproduced with either age or metallicity increasing
with increasing luminosity (or $\sigma$). However, if age were the
dominant effect, then the C-M relation would steepen with lookback
time (redshift), as the color of the young galaxies would get more
rapidly bluer compared to that of the old galaxies.  Instead, the
slope of the relation remains nearly the same (see Figure
8). Actually, the Kodama \& Arimoto argument can be applied also to
the color dispersion within a cluster, demonstrating that the
tightness of the C-M (and FP) relation in low-$z$ clusters cannot be
due to a conspiracy of age and metallicity being anticorrelated (as
advocated e.g., by Worthey, Trager \& Faber 1995). If so, the color
dispersion would rapidly increase with redshift, contrary to what is
seen in clusters up to $z\sim 1$ (see Figure 8).

\smallskip\pn
4.1.2 THE LUMINOSITY FUNCTION 
A cross check of the high formation
redshift of ETGs can be provided by looking at their luminosity 
in distant clusters. If ETGs evolve passively, following a pure
luminosity evolution, then their luminosities should increase
with increasing redshift by an amount that depends on the formation
redshift and on the slope of the IMF.

Initial attempts to detect the expected brightening of the
characteristic luminosity ($M^*$) of the luminosity function (LF) were
inconclusive, as Barger et al. (1998) failed to detect any appreciable
change between clusters at $z=0.31$ and $z=0.56$, possibly owing to
the small redshift baseline. On the other hand, comparing the LF of
$z\sim 0$ clusters to the LF of a sample of 8 clusters at
$0.40<z<0.48$ Barrientos \& Lilly (2003) found a brightening of the
characteristic luminosity $M^*$ consistent with passive evolution and
high formation redshift, also in agreement with the $(U-V)$ color
evolution of the red sequence. In a major cluster survey, De Propris
et al. (1999) explored the evolution of the observed $K$-band LF in 38
clusters with $0.1<z< 1$, and compared the results to the Coma
LF. With this much larger redshift baseline, De Propris et al. found
the trend of $K^*$ with redshift to be consistent with passive
evolution and $\zf\gsim 2$. They pointed out the agreement with the
results based on the color evolution of the red sequence galaxies, but
emphasized that this behavior of the LF implies that ``not only their
stellar population formed at high redshift, but that the assembly of
the galaxies themselves was largely complete by $z\sim 1$''. Kodama \&
Bower (2003) and Toft, Soucail, \& Hjorth (2003) came to the same
conclusions by studying the $K$-band LF of (respectively two and one)
clusters at $z\sim 1$. Breaking the $z=1$ barrier, Toft et al. (2004)
constructed a very deep $K$-band LF of a rich RDCS cluster at
$z=1.237$, and concluded that the most massive ellipticals that
dominate the top end of the LF were already in place in this
cluster. They compared the cluster $K$-band LF (corresponding to the
rest-frame $z$-band LF) to the $z$-band LF of local clusters (Popesso
et al. 2005) and derived a brightening by $\sim 1.4$ mag in the
rest-frame $z$-band characteristic magnitude, indeed as expected from
pure passive evolution.  Toft and colleagues also found a substantial
deficit of fainter ETGs, which could be seen as a manifestation of the
down-sizing effect in a high redshift cluster, a hint of which was
also noticed in other clusters at $z\sim 0.8$ (De Lucia et al. 2004).

However, a more complete study of 3 rich clusters at $1.1\lsim z\lsim
1.3$, including the $z=1.237$ cluster studied by Toft et al. (2004), did not
produce evidence of a down-sizing effect, down to at least 4
magnitudes below $K^*$ (Strazzullo et al. 2006). This study 
confirmed the brightening of $M_{\rm z}^*$ and  $M_{\rm K}^*$
by $\sim 1.3$ mag,
consistent with passive evolution of a population that formed at
$z\gsim 2$, and showed that the massive galaxies were already fully assembled
at $z\sim 1.2$, at least in the central regions of the 3
clusters. 

\smallskip\pn   
4.1.3 THE KORMENDY RELATION
An alternative way of detecting the expected brightening of old
stellar populations at high redshift is by tracing the evolution of
the Kormendy relation, which became relatively easy only after the
full image quality of HST was restored.  Thus, from HST data, the ETGs in
a cluster at $z=0.41$ were found brighter by $\Delta M_{\rm
K}=0.36\pm 0.14$ mag (Pahre, Djorgovski, \& de Carvalho 1996) or by
$0.64\pm 0.3$ mag (Barrientos, Schade, \& Lopez-Cruz 1996) with
respect to local galaxies, consistent (within such large errors) with
passive evolution of an old, single-burst stellar population. Schade
et al. (1996) using excellent-seeing CFHT (Canada-France-Hawaii Telescope)
imaging data for 3 clusters
at $z=0.23,\; 0.43$ and 0.55 detected a progressive brightening in
galaxy luminosity at a fixed effective radius that once more was
estimated to be consistent with passive evolution and formation at
high redshift. No differential evolution with respect to ETGs in the
cluster surrounding fields was detected.

Turning to HST data, a systematic brightening in the Kormendy relation, again 
consistent with passive evolution and high formation redshift,
was found by several other groups, eventually reaching redshifts $\sim 1$
(see Schade, Barrientos, \& Lopez-Cruz 1997, Barger et al. 1999, 
Ziegler et al. 1999, Holden et al. 2005a, Pasquali et al. 2006).

\smallskip\pn  
4.1.4 THE FUNDAMENTAL PLANE Besides the high spatial
resolution, constructing the FP of high redshift
cluster (and field) galaxies requires moderately high-resolution
spectroscopy (to get $\sigma$), hence a telescope
with large collective area. With one exception, for a few years this
was monopolized by the Keck Telescope, and FP studies of high-$z$
ellipticals first flourished at this observatory. In a crescendo
toward higher and higher redshifts, the FP was constructed for
clusters at $z=0.39$ (van Dokkum \& Franx 1996), $z=0.58$ (Kelson et
al. 1997), $z=0.83$ (van Dokkum et al. 1998, Wuyts et al. 2004 ), and
finally at $z=1.25$ and 1.27 (Holden et al. 2005b, van Dokkum \&
Stanford 2003). The early exception was the heroic study of two
clusters at $z=0.375$ using the 4m-class telescopes at ESO (NTT) and
Calar Alto (Bender, Saglia, \& Ziegler 1997, Bender, Ziegler, \&
Bruzual 1996, Bender et al. 1998).

\begin{figure}[t]
%\vskip-5.5truecm
\centerline{\psfig{file=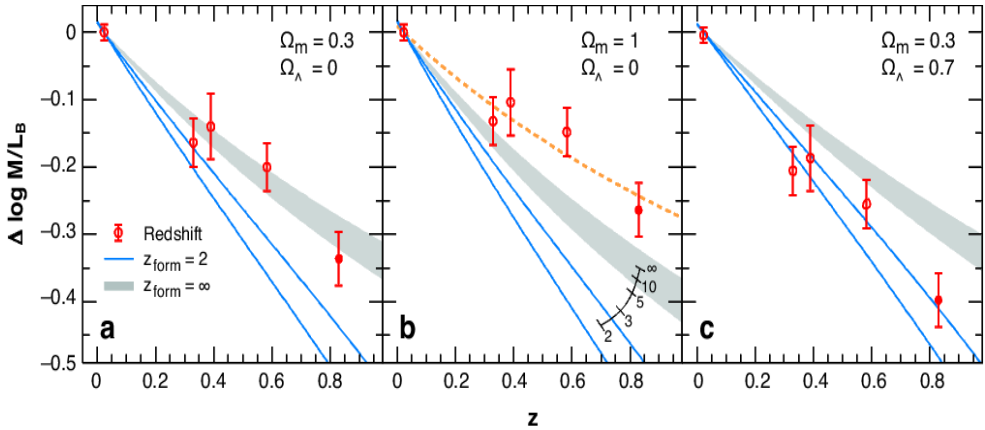,width=13cm,angle=-90,height=5.7cm}}

%\vskip-5truecm
%\plotone{feab.eps}
%\vskip 0.5truecm
\caption{\piedi The data points show the redshift evolution of the
 $M/\lb$ ratio of cluster elliptical galaxies as inferred from the
 shifts of the fundamental plane. The lines refer to the evolution of
 the $M/\lb$ ratio for stellar populations with a Salpeter initial mass function (IMF)
 ($s=2.35$) and formation redshifts as indicated in the left --and
 middle-- panel. The comparison is made for three different
 cosmologies. The dotted line in the middle panel shows a model with
 $\zf=\infty$ and a steep IMF ($s=3.35$).  From van Dokkum et
 al. 1998).  }
\end{figure}

\begin{figure}[t]
%\vskip-1.5truecm
\centerline{\psfig{file=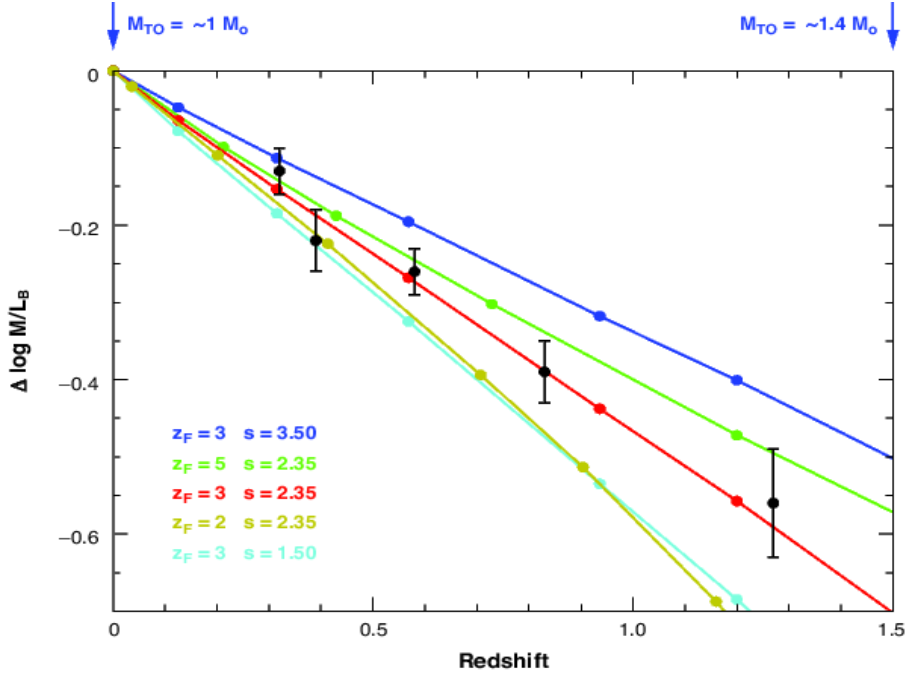,width=12cm,angle=-90,height=9cm}}

%\vskip-5truecm
%\plotone{feab.eps}
%\vskip-1truecm
\caption{\piedi The evolution with redshift of the $M_*/\lb$ ratio of
simple stellar populations of solar metallicity and various initial
mass function slopes ($dN\propto M^{-s}dM$) and formation redshifts,
as indicated.  The curves are normalized to their value at
$z=0$. Concordance cosmology ($\Omega_{\rm m}=0.3,\, \Omega_\Lambda
=0.7,\, H_\circ=70)$ is adopted.  The data points (from van Dokkum \&
Stanford 2003) refer to the shifts in the fundamental plane locations
for clusters at various redshifts. Note that for such high formation
redshifts the stellar mass at the main sequence turnoff is $\sim
1.4\,\msun$ at $z=1.5$ and $\sim 1.0\,\msun$ at $z=0$, as indicated by the
arrows. Adapted from Renzini (2005).  }
\end{figure}

The redshift evolution of the FP depends on a variety of factors. For
passive evolution  the FP shifts by amounts that depend on a
combination of IMF slope, formation redshift, and cosmological
parameters.  A systematic trend of the IMF slope with galaxy mass
would cause the FP to rotate with increasing redshift (Renzini \&
Ciotti 1993), as it would do for a similar trend in galaxy age.
An age dispersion ($\Delta t$) would cause the scatter
perpendicular to the FP to increase with redshift, as, for fixed $\Delta t$, 
$\Delta t/t$
increases for increasing redshift, i.e. decreasing galaxy age ($t$).
Clearly, the behavior of the FP with redshift can give a wealth of
precious information on the formation of ellipticals, their stellar
populations, and to some extent  on cosmology also (Bender et al. 1998).

All the quoted FP studies of high-$z$ clusters conclude that the FP actually
shifts nearly parallel to itself by an amount that increases with
redshift and is consistent with the passive evolution of stellar
populations that formed at high redshifts. The FP shifts imply a
decrease of the rest-frame $M/L$ ratio $\Delta\log M/\lb \simeq
-0.46z$ (van Dokkum \& Stanford 2003), but -- as emphasized above --
the formation redshift one can derive from it depends on both
cosmology and the IMF. Figure  9 illustrates the dependence on the
adopted cosmology. Note that for the ``old standard'' cosmology
($\Omega_{\rm M}=1$) galaxies would be older than the universe and
therefore the observed FP evolution can effectively rule out this
option (see also Bender et al. 1998).

Figure 10 shows the effect of the IMF slope and formation redshift on
the expected evolution of the $M/\lb$ ratio. The redshift range $z=0$
to 1.5 only probes the IMF between $\sim 1$ and $\sim 1.4\,\msun$,
which correspond to the masses at the main sequence turnoff $\mto$ of
oldest populations at $z=0$ (age $\sim 13$ Gyr, $\mto\simeq \msun$) and of
those at $z=1.5$ (age $\sim 4.5$ Gyr and $\mto\simeq 1.4\,\msun$).
Together, Figure 9 and 10 illustrate the formation
redshift/IMF/cosmology degeneracies in the FP diagnostics. However,
the cosmological parameters are now fixed by other observational
evidences, whereas the formation redshift as determined by the color
evolution of the cluster red sequence is independent of the IMF slope.
In summary, in the frame of the current concordance cosmology,
when combining the color and FP evolution of cluster ellipticals one
can conclude that the best evidence indicates a formation redshift
$\zf\sim 3$ and a Salpeter IMF slope in the pertinent stellar mass
range ($1<M<1.4\,\msun$).

The fact that the cluster FP does not appreciably rotate with increasing
redshift is documented down to $\sigma\sim 100\,\kms$ for clusters out
to $z\sim 0.3-0.4$ (lookback time $\sim 4$ Gyr) (Kelson et al. 2000,
van Dokkum \& Franx 1996) and down to $\sigma\sim 150\, \kms$ out to
$z\sim 0.8$ (lookback time $\sim 7$ Gyr) (Wuyts et al. 2004).  If the
large age trends with $\sigma$ derived in some of the studies using
the Lick/IDS indices were real (see Section 3.1), this would result in
a very large rotation of the FP in these clusters. For example, if age
were to increase along the FP from 5.5  Gyr to 13 Gyr (at $z=0$, and for
$\sigma=100$ and 320 \kms, respectively, see Nelan et al. 2005), then
at a lookback time of 4 Gyr ($z\sim 0.4$) the younger population would
have brightened by $\Delta M_{\rm B}\sim 1.33$ mag, and the older one
only by $\sim 0.46$ mag (using models from Maraston 2005), which results
in a FP rotation of $\sim 0.9$ mag in surface brightness.  
Alternatively, the much shallower age$-\sigma$ relation derived by Thomas
et al. (2005) implies an age increase from $\sim 9.5$ Gyr to $\sim 11.5$
Gyr for $\sigma$ increasing from 180 to 350 \kms, which implies a
rotation of the FP by $\sim 0.36$ mag in surface brightness by
$z=0.8$, which is still  consistent with the hint that in fact there may be
 a small FP rotation in a cluster at this redshift (Wuyts et al. 2004).

The scatter about the FP of clusters also remains virtually unchanged
with increasing redshift, however some of the claimed age-metallicity
anticorrelations derived from the Lick/IDS indices would result in a
dramatic increase of the scatter with redshift, causing the FP
itself to rapidly blur away.

\smallskip\pn  
4.1.5 THE LINE INDICES
The intermediate resolution spectra used for constructing the FP of
distant cluster ETGs, were also used to measure age-sensitive line
indices that can provide further constraints on the formation
epoch. Thus, Bender, Ziegler, \& Bruzual (1996) and Ziegler \& Bender
(1997) measured the $\mgb$ index of 16 ETGs in their two clusters at
$z=0.375$, and  found that the $\mgb -\sigma$ relation was shifted
toward lower values of the index. From such differences in $\mgb$
index, Ziegler \& Bender inferred that the age of the $z=0.375$
galaxies is about two thirds of that of ETGs in Coma and
Virgo. Therefore, $t(z=0)-\tf= 1.5\times [t(z=0.375)-\tf]$, where $t$
is the cosmic time and $\tf$ the cosmic time when the local and
distant cluster ETGs formed (which is assumed to be the
same). Adopting the $t-z$ relation for the concordance cosmology, one
derives $\tf\sim 1.7$ Gyr, corresponding to $\zf\gsim 3$.

From the strength of the Balmer absorption lines (H$_\delta$ and
H$_\gamma$) as age indicators, Kelson et al. (2001) used data for
several clusters up to $z=0.83$ and were able to set a lower limit to the 
formation redshift $\zf\gsim 2.5$, consistent with the above result from the 
$\mgb$ index.

In summary, the study of the stellar populations in ETGs belonging to
distant clusters up to $z\sim 1.3$ have unambiguously shown that these
objects have evolved passively from at least $z\sim 2-3$. This came
from the color, line strength, and luminosity evolution. Moreover, the
brightest cluster members at $z\sim 1-1.3$ and the characteristic
luminosity of the LF appear to be brighter than their local counterpart
by an amount that is fully consistent with pure passive evolution,
indicating that these galaxies were already fully assembled at this high 
redshift. This may not have been the case for less massive galaxies,
as their counts may be affected by incompleteness.

\smallskip\pn  
4.1.6 CAVEATS
Although it is well established that ETGs in
distant clusters are progenitors to their local analogs and formed at
high redshift, some caveats are nevertheless in order. First, as
frequently emphasized, the evidence summarized above only proves that
at least some cluster galaxies evolved passively from $z\gsim 1$
to the present, but other local ETGs may have $z\sim 1$ progenitors
that would not qualify as ETGs at that redshift, either
morphologically or photometrically. This ``progenitor bias'' (e.g.,
van Dokkum \& Franx 1996) would therefore
prevent us from identifying all the $z\sim 1$ progenitors of
local cluster ETGs, some of which may well be still star
forming. Second, the slope of the FP is progressively less accurately
constrained in higher and higher redshift clusters, because the
central velocity dispersion has been measured only for very few
cluster members (generally the brightest ones). Third, it is always
worth recalling that all luminosity-weighted ages tend to be biased
toward lower values by even minor late episodes of star formation.
Last, stellar population dating alone only shows when stars were formed,
not when the galaxy itself was assembled and reached its observed mass.

\subsection{Field versus Cluster Ellipticals up to $z\sim 1$}

\smallskip\pn 
In the local universe field ellipticals show small, yet
detectable differences compared to their cluster counterpart, being
possibly $\sim 1$ Gyr younger, on average. This $\Delta t$ difference,
if real, should magnify in relative terms and become more readily
apparent when moving to high redshift ($\Delta t/t$ is increasing).
Using the color evolution of the red sequence and the shift of the FP
with redshift, progress in investigating high$-z$ ETGs in the field
has been dramatic in recent years, along with the cluster versus field
comparison.

Schade et al. (1999) selected ETGs by morphology from the
Canada-France Redshift Survey (CFRS, Lilly et al. 1995a,b) and 
Low Dispersion Survey Spectrograph (LDSS) redshift survey (Ellis et
al. 1996), and constructed the rest-frame ($U-V$) C-M relation for
field ETGs in the $0.2<z<1.0$ range. They found that the C-M relation
becomes progressively bluer with redshift, with $\Delta(U-V)\simeq
-0.68\pm 0.11$ at $z=0.92$ with respect to the relation in Coma,
accompanied by a brightening by $\sim 1$ mag in the rest-frame $B$
band, as derived from the Kormendy relation. To be consistent with the
color evolution, this brightening should have been much larger than
observed if the color evolution were due entirely to the passive
evolution of stellar populations formed at high $z$.  Thus, Schade and
colleagues reconciled color and luminosity evolution by invoking a
residual amount of star formation (adding only $\sim 2.5\%$ of the
stellar mass from $z=1$ to the present), yet enough to produce the
observed fast color evolution. Support for such an interpretation
comes from about one third of the galaxies exhibiting weak [OII] emission,
which indicates that low-level star formation is indeed fairly widespread.
The rate of luminosity evolution was found to be identical to that of
cluster ellipticals at the same redshifts, hence no major
environmental effect was detected besides the mentioned low level of
star formation and a color dispersion slightly broader than in
clusters at the same redshift.

With COMBO-17, the major imaging survey project undertaken with the
ESO/MPG 2.2m telescope, Wolf et al. (2003) secured deep optical
imaging in 17 broad and intermediate bands over a total 0.78 square
degree area, from which Bell et al. (2004b) derived photometric
redshifts accurate to within $\delta z\sim 0.03$.  The bimodality of
the C-M relation, so evident at $z\sim 0$ (e.g., Baldry et al. 2004),
clearly persists all the way to $z\sim 1.1$ in the COMBO-17 data, and this
allowed Bell and colleagues  to isolate $\sim 5,000$ ``red sequence'' ETGs
down to $R<24$. As mentioned above, $\sim 85\%$ of such
color-selected galaxies appear also morphologically early-type on the
ACS images of the GEMS (Galaxy Evolution from Morphology and SED) 
survey (Rix et al. 2004, Bell et al. 2004a).
The rest-frame ($U-V$) color of ETGs in the COMBO-17 survey evolves by
a much smaller amount than that reported by Schade et al. (1999) for
the morphologically-selected ETGs, i.e., by only $\sim 0.4$ mag
between $z=0$ and 1, as expected for an old stellar population
that formed at high redshift ($\zf\gsim 2$). This color evolution is
also in agreement with the $\sim 1.3$ mag brightening of the
characteristic luminosity $M_{\rm B}^*$ in the Schechter fit to the
observed LF. Thus, when  comparing only  the color and $M_{\rm B}^*$
evolution, the field ETGs in the COMBO-17 sample seem to evolve in
much the same fashion as their cluster counterparts. Using COMBO-17
data and GEMS HST/ACS imaging, McIntosh et al. (2005) studied a sample
of 728 morphology- and color-selected ETGs, finding that up to
$z\sim 1$ the Kormendy relation evolves in a manner that is consistent
with the pure passive evolution of ancient stellar populations.

\smallskip\pn
\begin{figure}[]
%\vskip-1.5truecm
\centerline{\psfig{file=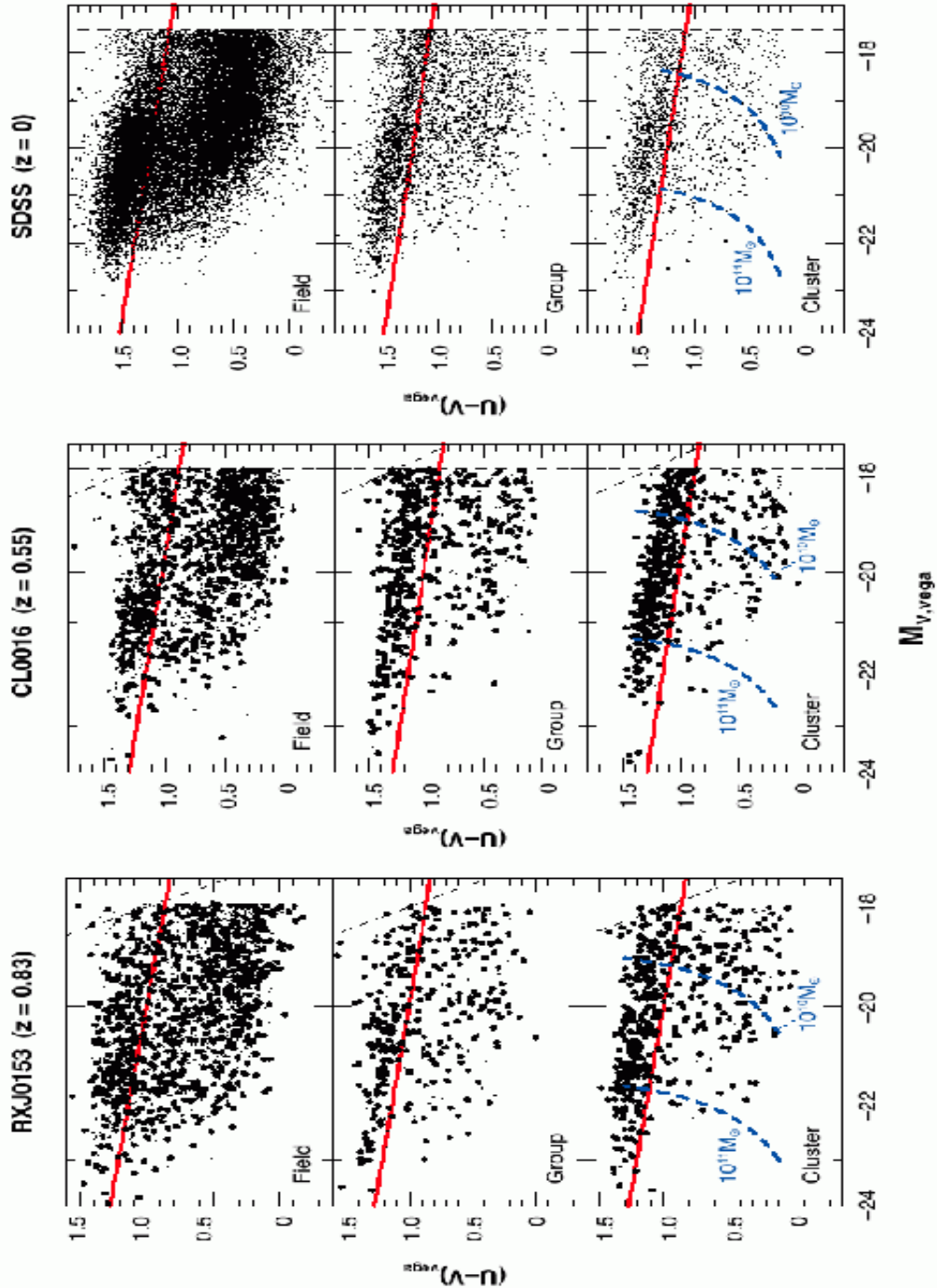,width=14cm,angle=0,height=18.5cm}}

%\vskip-5truecm
%\plotone{feab.eps}
%\vskip-1truecm
\caption{\piedi The C-M relations of two high redshift clusters and
their surrounding fields at progressive lower density (labelled
``group'' and ``field'') are compared to their local counterparts from
the Sloan Digital Sky Survey. The red line marks the adopted separation
between the red sequence galaxies, and the blue, star-forming
galaxies. In the cluster panels the blue dashed lines show the
approximate location of galaxies with stellar mass of $10^{10}$
(right) and $10^{11}\msun$ (left). From Tanaka et al. (2005).  
}
\end{figure}

From deep Subaru/Suprime-Cam imaging over a 1.2 deg$^2$ field [covered
by the Subaru-XMM Deep Survey (SXDS)], Kodama et al.  (2004) selected
ETGs for having the $R-z$ and $i-z$ colors in a narrow range as
expected for passively evolving galaxies in the range $0.9\lsim z\lsim
1.1$, and the sampled population included both field and cluster
ETGs. They found a deficit of red galaxies in the C-M sequence $\sim
2$ mag fainter than the characteristic magnitude (corresponding
to stellar masses below $\sim 10^{10}\msun$).  Less massive galaxies
appear to be still actively star-forming, while above $\sim 8\times
10^{10}\msun$ galaxies are predominantly passively evolving. This was
interpreted as evidence for down-sizing in galaxy formation (\`a la
Cowie et al. 1996), with massive galaxies having experienced most of
their star formation at early times and being passive by $z\sim 1$,
and many among the less massive galaxies experience extended star
formation histories.

In a comprehensive study also based on deep, wide-field imaging with
the Suprime-Cam at the Subaru Telescope, Tanaka et al. (2005) obtained
photometric redshifts based on four or five optical bands and
constructed the C-M relations for the two clusters (at $z=0.55$ and
0.83) included in the field, and for their extended
environment. Tanaka and colleagues further distinguished galaxies in
the cluster environment as belonging either to recognized ``groups'',
or otherwise to the ``field''.  The results are shown in Figure 11,
where the various plots allow one to visually explore trends with
redshift for given environment, or with environment for given
redshift. The red sequence appears already in place in the ``field''
in the highest redshift sample, but no clear color bimodality is
apparent. (Note that COMBO-17 does find bimodality at this redshift,
possibly due to its photometric redshifts based on many more bands
being more accurate.)  The color bimodality is instead clearly
recognizable in the $z=0.55$ ``field'' sample. At higher environmental
densities (labelled ``group'') the C-M relation of the red sequence is
clearly recognizable already in the $z=0.83$ sample, and even more so
in the ``cluster'' sample. Tanaka and colleagues argue that the bright
and the faint ends of the red sequence are populated at a different
pace in all three environments; the more massive red galaxies are
assembled first, i.e., the C-M relation grows from the bright end to
the faint end in all three environments (not in the opposite way, as
one may naively expect in a hierarchical scenario), which is
interpreted in terms of down-sizing. Note also that the faint end
appears to be well in place in ``clusters'' at $z=0$ while virtually
still lacking in the field (see also Popesso et al. 2005). Using
HST/WFPC2 imaging over a $\sim 30$ arcmin$^2$ field including the same
$z=0.83$ cluster, Koo et al.  (2005) were able to measure the
rest-frame $(U-B)$ colors of the sole bulge component of 92 galaxies
with $M_{\rm B}<-19.5$, part in the cluster itself, part in the
surrounding field. Their very red color does not show any
environmental dependence, suggesting similarly old ages and high
formation redshifts.

The public delivery of the Hubble Deep Field data (HDF, Williams et
al. 1996) spurred several studies of field ETGs. Thus, Fasano et
al. (1998) applied the Kormendy relation to a sample of
morphologically selected ETGs in HDF-North and estimated an increase
of the surface brightness at a fixed effective radius that was
consistent with a high formation redshift ($\zf\sim 5$), according to
the galaxy models by Bressan, Chiosi \& Fagotto (1994) and Tantalo et
al. (1996).

Although the LF and the C-M and Kormendy relations had already given
useful indications on the analogies and differences between cluster
and field ETGs, major progress came with the study of the differential
evolution (field versus cluster) of the FP with redshift. In early
studies, no field/cluster difference had clearly emerged at $z=0.3$
(Treu et al. 1999), $z\sim 0.4$ (Treu et al. 2001), $z=0.55$ (van
Dokkum et al 2001), and $z=0.66$ (Treu et al. 2002). But already at
these modest redshifts there were hints that the brightest, most
massive ETGs in the field closely follow the FP evolution of their
cluster counterparts, while less massive ETGs (especially S0's) appear
to evolve slightly faster, and hence look younger. This was more
accurately quantified for morphologically-selected ETGs up to $z\sim
1$ in the HDF-North by van Dokkum \& Ellis (2003), showing a field
versus cluster difference $\Delta {\rm ln} M/\lb= -0.14\pm 0.13$ in the
FP. This implies that field ETGs are on average younger by only
$16\%\pm15\%$ at $<\!z\!>=0.88$. Van Dokkum \& Ellis also inferred
that the bulk of stellar mass in the observed ETGs must have formed at
$z\gsim 2$ even in the field, with only minor star formation at lower
redshifts. Then, moving to the wider GOODS (Great Observatories Origin
Deep Survey)-South field (Giavalisco et al.  2004a), van der Wel et
al. (2004, 2005a) constructed the FP for a total of 33 color and
morphology-selected ETGs at $0.60<z<1.15$, using intermediate
resolution spectra taken at the ESO Very Large Telescope (VLT). They
also found the most massive galaxies ($M_*>2\times 10^{11}\msun$) to
behave much like their cluster analogs at the same redshifts, while
less massive galaxies appeared to be substantially younger.  Moreover,
all these studies noted the higher proportion of weak [OII] emitters
among the field ETGs ($\sim 20\%$) compared to their cluster
counterparts, as well as the higher proportion of galaxies with strong
Balmer lines (the K+A Galaxies).

\begin{figure}[t]
%\vskip-2.5truecm
\centerline{\psfig{file=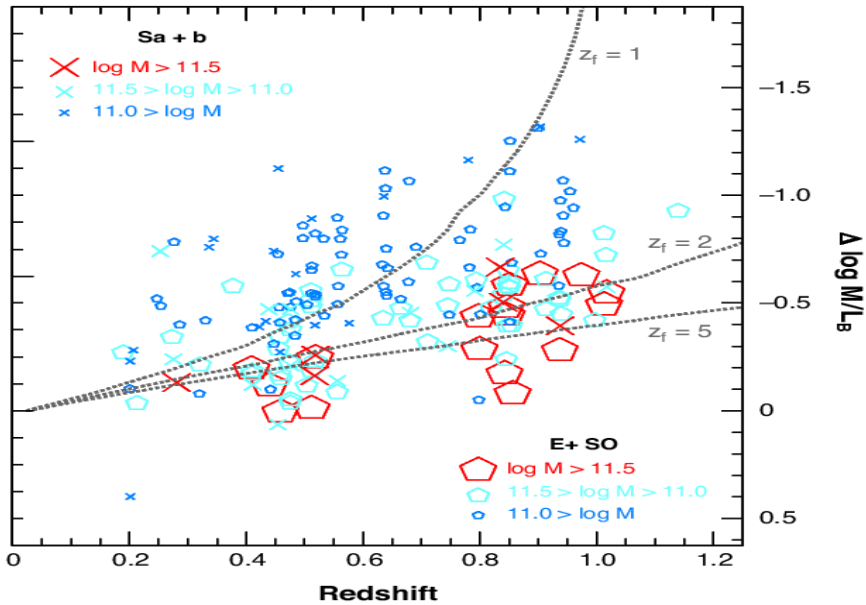,width=11.5cm,angle=0,height=8cm}}

%\vskip-5truecm
%\plotone{feab.eps}
%\vskip-1truecm
\caption{\piedi The offset $\Delta\log M/\lb$ from the fundamental
plane of cluster ellipticals at $z\sim 0$ for the early-type galaxies
in the GOODS-North field (from Treu et al. 2005b). Different symbols
are used for early-type (E+S0) galaxies and bulges in late-type
(Sa+Sb) galaxies, as well as for the various stellar mass ranges as
indicated. The dotted lines are labelled by various formation
redshifts.  
}
%\vskip-8truemm
\end{figure}

The main limitations of all these early studies was in the small
number of objects observed at each redshift, which must go a long way
toward accounting for occasional discrepancies in the results.  In a
major effort to overcome this limitation, Treu et al. (2005a,b)
obtained high-resolution spectra at the Keck telescope for 163
morphologically-selected ETGs in the GOODS-North field, which were
distributed over the redshift range $0.2<z<1.2$.  The main results of
this study are displayed in Figure 12, showing that the most massive
ellipticals in the field do not differ appreciably from their cluster
analogs in having luminosity-weighted ages implying $\zf\gsim
3$. However, the lower the mass the larger the dispersion in the
$M/\lb$ ratio, with a definite trend toward lower values with
decreasing mass, implying lower and lower formation redshift. This
demonstrates that completion of star formation in field galaxies
proceeds from the most massive to the less massive ones, as is indeed
expected from the down-sizing effect (Cowie et al. 1996, Kodama et
al. 2004) and is consistent with the scenario shown in Figure 6.  This
systematic trend in the $M/L$ ratio with galaxy mass results in a
``rotation'' of the FP with increasing redshift, as less massive
galaxies evolve faster in luminosity compared to the more massive,
older ones (but one should beware of possible Malmquist bias). This
result confirms early hints for a modest rotation of the FP of field
ETGs with redshift, as also does a study of the FP of 15 ETGs at
$0.9\lsim z \lsim 1.3$ by di Serego Alighieri et al. (2005), a sample
drawn from the K20 survey (Cimatti et al. 2002b).  Figure 13 shows the
FP for the combined Treu et al.  (2005b) and di Serego Alighieri et
al. (2005) samples of ETGs with $<\! z\!>=1.1$, where the rotation
with respect to the Coma FP is apparent. Note that a similar FP
rotation in two clusters at $z=0.83$ and 0.89 has been recently
unambiguously detected by J{\o}rgensen et al. (2006), having extended
the $\sigma$ measurements below $\sim 100\;\kms$. Somewhat at variance
with the FP studies reported above was the Deep Groth Strip Survey
result (Gebhardt et al. 2003), in which no difference in the slope was
found up to $z\sim 1$ compared to the local FP (hence no down-sizing),
which was coupled to a much faster luminosity evolution compared to
all other results. Treu et al. (2005b) discuss the possible origins of
the discrepancy, and attribute it to a combination of selection bias,
small number statistics, and relatively low S/N spectra.

From population synthesis models one expects the rate of evolution of
the $M/L$ ratio to be slower at longer wavelengths compared to the $B$
band, because it is less affected by the main sequence turnoff moving
to cooler temperatures with increasing age. This expectation was
qualitatively confirmed by van der Wel et al. (2005b), who found
$\Delta{\rm ln}\,(M/\lb)=-(1.46\pm 0.09)z$ and $\Delta{\rm
ln}\,(M/\lk)=-(1.18\pm 0.1)z$, which appears to be in agreement with
the prediction of some models (Maraston 2005), but not of others
(Bruzual \& Charlot 2003), possibly owing to the different treatment of
the AGB contribution.

Instead of measuring central velocity dispersions directly, Kochanek et al. 
(2000) and Rusing and Kochanek (2005) estimated them from the lens geometry 
for a sample of (field) lensing ETGs at $0.2\lsim z\lsim 1$. The resulting FP 
shifts appear to be similar to those of cluster ETGs, although with more 
scatter, and indicate $<\!\zf\!>\gsim 1.5$ for the bulk of the stars in
the lensing galaxies.

In summary, like at low redshifts, also at $z\sim 1$ there appear to be small
detectable differences between ETGs in high- and low-density regions, but such 
differences are more evident for faint/low-mass galaxies than for the bright
ones.

\begin{figure}[t]
%\vskip-0.5truecm
\centerline{\psfig{file=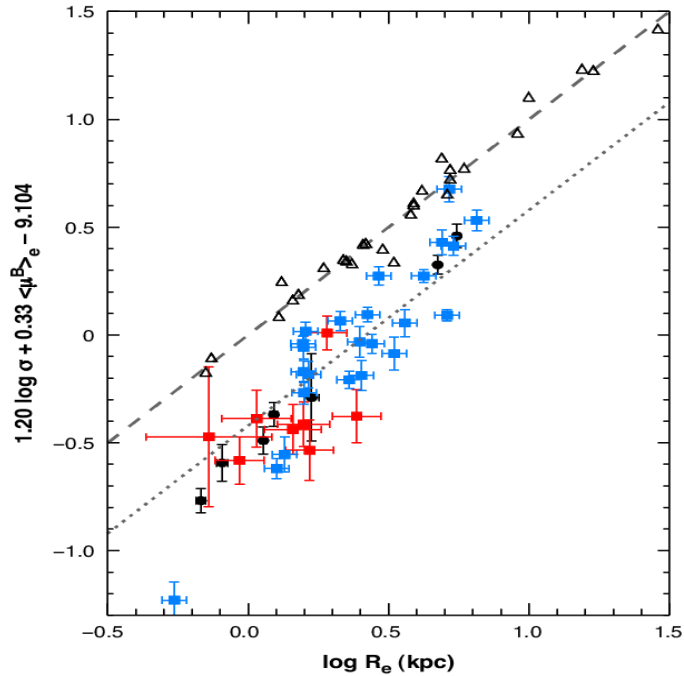,width=9cm,angle=0,height=9cm}}

%\vskip-5truecm
%\plotone{feab.eps}
%\vskip 1truecm
\caption{\piedi An edge-on view of the fundamental plane for field
ETGs at $z\sim 1.1$ from di Serego Alighieri et al. (2005, red squares
and black circles) and Treu et al. (2005b, green squares). Open
triangles refer to the Coma ellipticals from J{\o}rgensen, Franx, \&
Kj{\ae}rgaard (1995) and the dashed line is a best fit to the
data. The dotted line is shifted parallel to the dashed line by an
amount in surface brightness corresponding to the observed shift of
the fundamental plane of galaxy clusters (i.e., $\Delta M/\lb=-0.46z$, 
van Dokkum \&
Stanford 2003). The effective surface brightness in the $B$ band
($\mu^{\rm B}_{\rm e}$) is in magnitudes per arcsec$^2$.  
}
%\vskip-0.8truecm
\end{figure}

\subsection{Ellipticals Beyond $z\sim 1.3$}

\par
\noindent
Up to $z\sim 1.3$ the strongest features in the optical spectrum of
ETGs are the CaII H\&K lines and the 4000 \AA\ break. But at higher
redshifts these features first become contaminated by OH atmospheric
lines, and then move to the near-IR, out of reach of CCD
detectors. The lack of efficient near-IR multi-object spectrographs
(even in just the $J$ band) has greatly delayed the mapping of the ETG
population beyond $z\sim 1.3$. Thus, for almost a decade the most
distant spectroscopically confirmed old spheroid was an object at
$z=1.55$ selected for being a radiogalaxy (Dunlop et al. 1996, Spinrad
et al. 1997). The spectral features that made the identification
possible included a set of FeII, MgII and MgI lines in the rest-frame
near-UV, in the range of $\sim 2580-2850$ \AA, which is typical of
F-type stars.  The UV Fe-Mg feature offers at once both the
opportunity to measure the redshift, and to age-date the galaxy,
because it appears only in populations that have been passively
evolving since at least a few $10^8$ years. It has been also used to
age-date local ETGs (e.g., Buson et al. 2004). Thus, using this
feature, Spinrad and colleagues inferred an age of $\sim 3.5$ Gyr, implying
$\zf>5$ (even in modern cosmology).

This record for the highest redshift ETG was eventually broken by
Glazebrook et al.  (2004) and Cimatti et al. (2004), using the
same features in the rest-frame UV (see Figure 14). They reported the
discovery of, respectively, five passively evolving galaxies at
$1.57\lsim z\lsim 1.85$, and four other such objects at $1.6\lsim z\lsim
1.9$. All being  brighter than $K=20$, these galaxies are quite
massive ($M\gsim 10^{11}\msun$), and hence would rank among the most
massive galaxies even in the local universe. This suggests that they
were (almost) fully assembled already at this early epoch, and having been 
passive since at least $\sim 1.1$ Gyr had to form
at redshift $\gsim 2.7$ . The four objects found by Cimatti and colleagues
are included in the GOODS-South field, and the GOODS deep HST/ACS imaging
showed that two objects are definitely elliptical galaxies, and the two
others are likely to be S0's.

Though breaking the old redshift record was certainly an exciting
result, perhaps far more important was the discovery that the surface
density of $z>1.5$ ETGs is indeed much higher than one would have
expected from just the single object found by Dunlop et
al. (1996). Indeed, this galaxy was selected from a catalog of
radiogalaxies covering a major fraction of the whole sky, whereas the
nine galaxies in the Cimatti and colleagues and Glazebrook and
colleagues samples come from a combined area of only 62 arcmin$^2$.

\begin{figure}[t]
%\vskip-2.5truecm
\centerline{\psfig{file=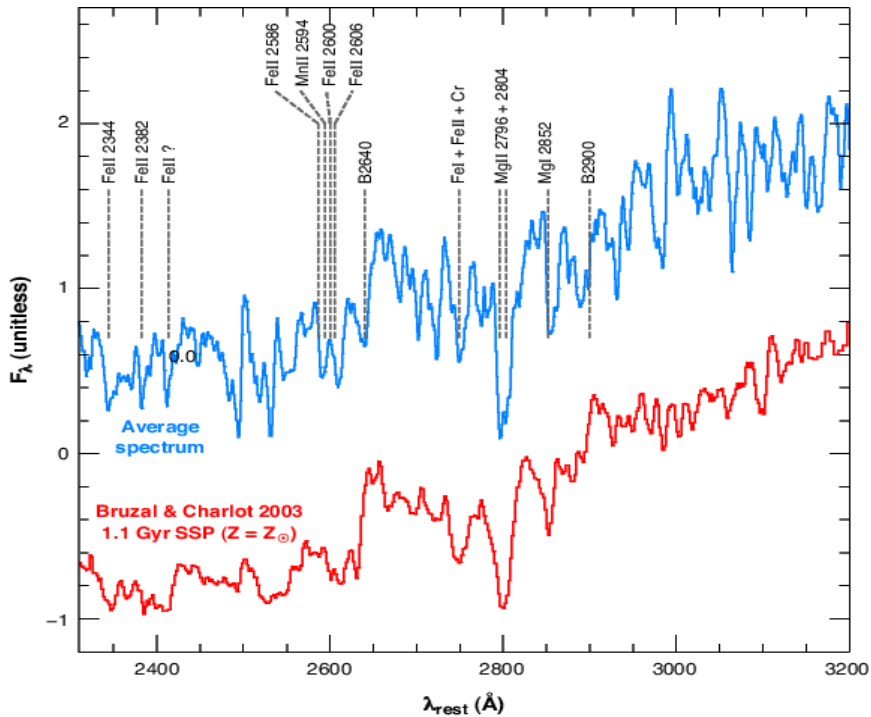,width=11.5cm,angle=0,height=9.5cm}}

%\vskip-5truecm
%\plotone{feab.eps}
%\vskip-1truecm
\caption{\piedi The rest-frame coadded spectrum of the four passively
evolving galaxies at $1.6< z< 1.9$ with the identification of the main
spectral features (blue spectrum). The spectrum from Bruzual \&
Charlot (2003) for a 1-Gyr-old SSP (simple stellar population) model
of solar metallicity is also shown (red spectrum). (From Cimatti et
al. 2004) 
}
\end{figure}

Further identifications of very high redshift ETGs used this UV
feature: McCarthy et al. (2004) reported the discovery of 20 ETGs with
$1.3\lsim z\lsim 2.15$ and $K<20$ as part of the Gemini Deep Deep
Survey (GDDS) (including the 5 galaxies from Glazebrook et al.
2004). Within the $\sim 11$ arcmin$^2$ area of the Hubble Ultra Deep
Field (HUDF, Beckwith et al. in preparation), Daddi et al. (2005b) identified 7
ETGs with $1.39\lsim z\lsim 2.5$ using their HST/ACS grism
spectra. For all these objects the stellar mass derived from the
spectral energy distribution (SED, typically extending from the $B$ to
the $K$ band) is in excess of $\sim 10^{11}\msun$. Over the same
field, Yan et al.  (2004) identify 17 objects with photometric
redshifts between 1.6 and 2.9, whose SED can be best fit by a dominant
$\sim 2$ Gyr old stellar population, superimposed to a low level of
ongoing star formation.

Rather than digging deep into small fields, Saracco et al. (2005)
searched for bright high-$z$ ETGs over the $\sim 160$ arcmin$^2$ field
of the MUNICS survey (Drory et al. 2001), and selected objects with
$R-K>5.3$ and $K<18.5$ for spectroscopic follow up with the
low-resolution, near-IR spectrograph on the TNG 3.5m telescope.  They
identified 7 ETGs at $1.3\lsim z\lsim 1.7$, all with mass well in
excess of $10^{11}\msun$.

Altogether, these are virtually all the spectroscopically confirmed
ETGs at $z>1.3$ known to date. Given the long integration time needed
to get spectroscopic redshifts, it became clear that an effective
criterion was indispensable to select high-$z$ ETG candidates. To this
end, Daddi et al.  (2004) introduced a robust criterion based on the
$B-z$ and $z-K$ colors, that very effectively selects galaxies at
$1.4\lsim z\lsim 2.5$ (the so-called BzKs), and among them separates
the star forming BzKs with $BzK\equiv (z-K)_{\rm AB} -(B-z)_{\rm
AB}\ge -0.2$, from the passive ones, with $BzK < -0.2$ and $(z-K)_{\rm
AB}>2.5$. The criterion is primarily an empirical one, based on the
spectroscopic redshifts from the K20 survey (Cimatti et al. 2002b) and
other publicly available data sets. However, Daddi and colleagues
showed that synthetic stellar populations of the two kinds (i.e., star
forming and passive) do indeed occupy the corresponding areas in this
plot, when redshifted to $1.4<z<2.5$.  An application of the method to
a 320 arcmin$^2$ field is shown in Figure 15 (Kong et al. 2006). In
this latter study, it is estimated that the space density of massive
and passive BzKs (with $K<20$, stellar mass $\gsim 10^{11}\msun$ and $<\!
z\!>\simeq 1.7$) is $20\pm 7\%$ that of $z=0$ ETGs within the same
mass limit. Then there appears to be a sharp drop of passive BzKs
beyond $z=2$, which in part may be due to the available $B$-band data being
not deep enough (Reddy et al. 2005)

Further candidate ETGs with masses up to a few $10^{11}\msun$ have
been identified at even higher redshifts, such as six objects within
the HUDF at a (photometric) redshift $>2.8$ (Chen \& Marzke 2004),
where both the redshift and the old age are inferred from the observed
break between the $J$ (F110W) and the $H$ (F160W) band being interpreted as
the 4000 \AA\ break.  One of these objects is undetected in the deep
GOODS optical data, but is prominent in the GOODS Spitzer/IRAC $3.5-8\;
\mu$m images (M. Dickinson et al., in preparation). Thus 
its SED shows two breaks, one between the $z$ and the $J$ band, and one
between the $K$ and the 3.5 $\mu$m IRAC band. Identifying them
respectively with the Lyman and Balmer breaks, the object would be placed at
$z\sim 6.5$, it would be passively evolving with $\zf >9$, and would have the
uncomfortably large mass of a few $10^{11}\msun$ (Mobasher et
al. 2005). Lower redshift alternatives give much worse fits to
the data, whereas the use of models with strong AGB contribution 
(Maraston 2005) results in a somewhat less extreme mass and formation redshift.

\subsection{Evolution of the Number Density of ETGs to $z\sim 1$ and Beyond}

The studies illustrated so far have shown that ETGs exist up to $z\sim
1$, both in clusters and in the field, and are dominated by old
stellar populations that formed at $z\gsim 2-3$. Moreover, a handful
of ETGs has also been identified (over small fields) well beyond
$z\sim 1$. Some of these ETGs appear to be as massive as the most
massive ETGs in the local universe, demonstrating that at least some
very massive ETGs are already fully assembled at $z\gsim 1$. However,
the expectation is for the number of ETGs to start dropping at some
redshift, when indeed entering into the star formation phase of these
galaxies, or when they were not fully assembled yet.  Therefore, what
remained to be mapped by direct observations was the evolution with
redshift of the comoving number density of ETGs, and to do so as a
function of mass and environment while covering wide enough areas of
the sky in order to reduce the bias from cosmic variance. Only in this
way one could really overcome the so-called progenitor
bias. Because deep and wide surveys require so much
telescope time, progress has
been slow. Cosmic variance may still be responsible for the apparent
discrepancies between galaxy counts from different surveys, but
occasionally the interpretation itself of the counts may be prone to
ambiguities.

One of the main results of the CFRS was that the number density of red
galaxies shows very little evolution over the redshift range $0<z<1$ (Lilly et
al.  1995b). Following this study, in an attempt to map the number evolution 
of ETGs all the
way to $z\sim 1$ , Kauffmann, Charlot, \& White (1996) extracted 90
color-selected ETGs without [OII] emission from the CFRS redshift
catalog. They used a $V/V_{\rm max}$ test and concluded that at $z=1$
only $\sim 1/3$ of bright ETGs had already assembled or had the colors
expected for old, pure passively evolving galaxies. However,
Im et al. (1996) identified $\sim 360$ ETGs morphologically selected
on archival HST images, and also conducted the $V/V_{\rm max}$ test using
photometric redshifts, finding no appreciable number density evolution
up to $z\sim 1$ and a brightening consistent with passive
evolution. The $V/V_{\rm max}$ test was repeated --  again using the
CFRS sample -- by Totani \& Yoshii (1998) who concluded that there was
no evolution in the number density up to $z\sim 0.8$, and ascribed
 the apparent drop at $z>0.8$ to a color selection bias. No evolution
of the space density of morphologically-selected ETGs up to $z\sim 1$
was found by Schade et al. (1999) too, who used the HST imaging of the
CFRS and LDSS redshift surveys.

\begin{figure}[t]
%\vskip-2.5truecm
\centerline{\psfig{file=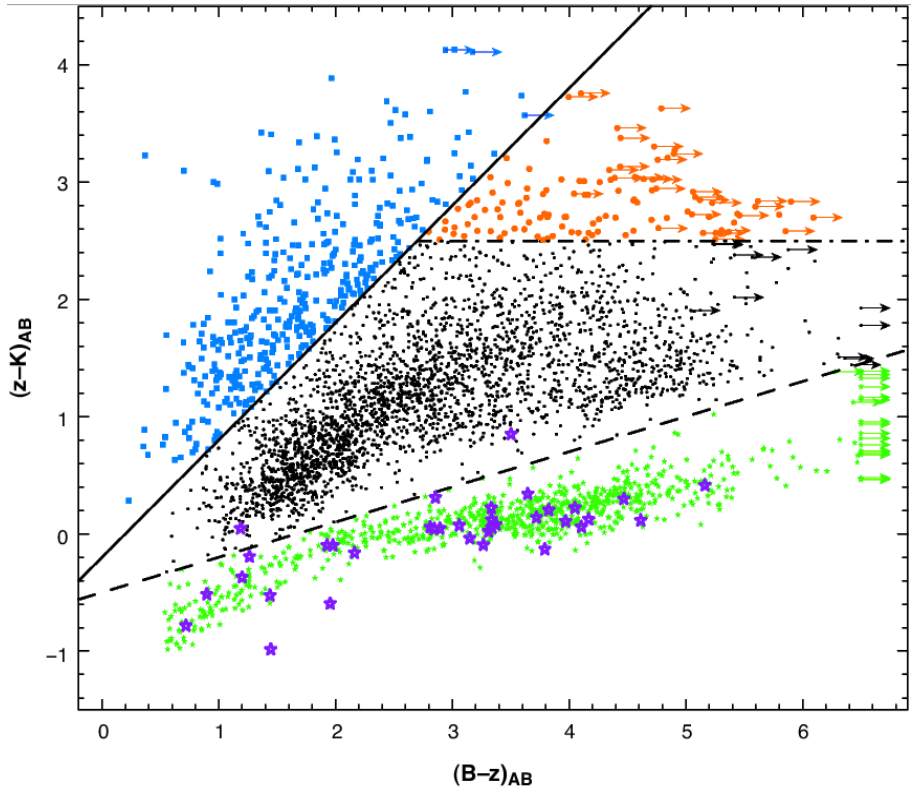,width=12.0cm,angle=-90,height=10.5cm}}
\caption{\piedi The $BzK$ plot introduced by Daddi et al. (2004) is
here shown for objects to a limiting magnitude $K_{\rm Vega}=20$ from
a 320 arcmin$^2$ field (from Kong et al. 2006). Black dots refer to galaxies
at $z<\sim 1.4$, blue dots refer to starforming galaxies at $\sim 1.4 <
z<\sim 2.5$, orange dots refer to passively evolving
galaxies at $\sim 1.4 < z< \sim 2.5$. Green dots are
Galactic stars in the same field and purple stars are local stellar
standards.
}
\end{figure}

Several attempts to trace the evolution of the number density of
morphologically-selected ETGs to the highest possible redshifts were made
using HDF data.  In some of these studies very little, if any,
change in the space density of ETGs was found up to $z\sim 1$ (e.g.,
Driver et al. 1998, Franceschini et al. 1998, Im et al. 1999), or up
to even higher redshifts when combining the HDF optical data with very deep
near-IR data (Benitez et al 1999, Broadhurst \& Bouwens 2000). These
latter authors emphasized that without deep near-IR data many high-$z$
ETGs are bound to remain undetected, and that spectroscopic
incompleteness beyond $z\sim 0.8$ is partly responsible for some of
the previous discrepancies.  Beyond $z\sim 1$ a drop in
the space density of ETGs was detected in several studies including
the HDF (Zepf 1997, Franceschini et al. 1998, Barger et al. 1999,
Rodighiero, Franceschini, \& Fasano 2001, Stanford et al. 2004). In
particular, Stanford and colleagues applied the $V/V_{\rm max}$ test to a
sample of 34 ETGs from the HDF-North that includes deep NICMOS imaging
in the $H$ band (F160W), and concluded for a real drop at $z>1$, but
advocated the necessity to explore much wider fields in order to
improve the statistics and cope with cosmic variance. Finally, the
HDF-South optical data were complemented by ultra-deep $JHK$ imaging
at the VLT (the FIRES survey, Labb\'e et al.  2003, Franx et
al. 2003), revealing a population of near-IR galaxies with very red
colors ($J-K>2.3$), called distant red galaxies (DRG), a fraction of
which may be ETGs at very high redshifts (see below). Compared to
HDF-North, its Southern equivalent appeared to be much richer in very
red galaxies, e.g., including 7 objects with $(V-H)_{\rm AB}>3$ and
$H_{\rm AB}<25$ while HDF-North has only one.  Clearly,
exploring much wider fields compared to HDF's $\sim 5$ arcmin$^2$
field was imperative in order to make any significant progress.

Passive ETGs formed at very high redshift (e.g., $z>3$) would indeed
have very red colors at $z\gsim 1$, and thus they should be found
among the so-called extremely red objects (ERO), a class defined for
having $R-K>5$ (or similar color cut), and whose characteristics and
relation to ETGs have been thoroughly reviewed by McCarthy
(2004). Using a much shallower sample than that from the HDF, but one
that covers an area $\sim 140$ times wider than it, Daddi et
al. (2000) and Firth et al. (2002) showed that EROs are much more
abundant than previously found in smaller fields and are much more
strongly clustered than generic galaxies to the same limiting
magnitude $K\sim 19$. This made them likely candidates for high-$z$
ETGs, and assuming that $\sim 70\%$ of EROs are indeed ETGs at $z>1$,
Daddi, Cimatti \& Renzini (2000) concluded that most field ellipticals
were fully assembled by $z\sim 1$. However, Cimatti et al. (2002a)
actually found that out of the 30 EROs with secure redshifts and
$K<19.2$, only $50\%$ are passively evolving objects and these are
distributed in the redshift interval $0.8\lsim z\lsim 1.3$, while the
other 50\% is made by highly-reddened, actively star-forming
galaxies. Interestingly, precisely 50\% among a sample of 129 EROs
with $K<20.2$ have been detected at 24 $\mu$m with Spitzer/MIPS,
reinforcing the conclusion that up to one half of EROs are likely to
be passive precursors to ETGs (Yan et al. 2004). In fact, the fraction
of passive EROs decreases to $\sim 35\%$ on a spectroscopic complete
sample to $K=20$ (Cimatti et al. 2003).  Nevertheless, the number
density of passive EROs appeared to be broadly consistent with no
density evolution of ETGs up to $z\sim 1$, or a modest decrease.

With the COMBO-17 survey Bell et al. (2004b) went a long way toward
coping with cosmic variance. With their 5,000 color-selected ETGs up
to photometric redshift $z\sim 1.1$, Bell and colleagues were able to
construct their rest-frame $B$-band luminosity functions in nine
redshift bins ($0.2<z <1.1$), and derived the best fit Schechter
parameters for them using a fixed value of the faint-end slope,
$\alpha=-0.6$.  They found that the characteristic luminosity $M_{\rm
B}^*$ brightens by $\sim 1.0$ mag between $z=0.25$ and 1.05,
consistent with passive evolution within the errors, and also with the
brightening expected from the FP shift ($\Delta\log\, M/\lb =
-0.46\Delta z$), which predicts $\sim 0.9$ mag. At the same time, the
normalization factor $\phi^*$ drops by a factor of $\sim 4$, but much
of the drop is in the highest redshift bins which may be affected by
incompleteness. More robust than either $\phi^*$ or $L^*$ separately,
is their product $\phi^*L_{\rm B}^*$ which is proportional to the
$B$-band luminosity density, and this is found to be nearly constant
up to $z\sim 0.8$. This is at variance with a pure passive-evolution
scenario, that would have predicted an increase by a factor of $\sim
2$.  Thus, the color of the COMBO-17 red sequence follows nicely the
expectation from passive evolution (cf. section 4.2), but the number
density of red sequence galaxies does not, and Bell and colleagues
concluded that the stellar mass in red sequence galaxies has nearly
doubled since $z\sim 1$.

In a major observational effort at the Keck telescope, Faber et
al. (2006, DEEP2 project) secured spectroscopic redshifts for $\sim
11,000$ galaxies with $R<24.1$, and also reanalyzed the COMBO-17 data,
finding separate best fit Schechter parameters in various redshift
bins up to $z\sim 1.1$. Faber and colleagues emphasize that $\phi^*$ and $M^*$
are partly degenerate in these fits, for which the faint-end slope was
fixed at $\alpha=-0.5$. Thus, between $z=0.3$ and 1.1, $M^*$ brightens
by $\sim 0.47$ mag and $\phi^*$ drops by a factor of $\sim 2.5$ for the
DEEP2 data, and respectively up by $\sim 0.95$ mag and down by a
factor of $\sim 4$ for the COMBO-17 data. Once more, much of the $\phi^*$
drops are confined to the last redshift bin, and emphasis is placed on
both DEEP2 and COMBO-17 confirming that the $B$-band luminosity
density is nearly constant up to $z\sim 0.8$, along with the
implication that the mass density in ETGs has increased, presumably by
a factor of $\sim 2$, as estimated by Bell et al. (2004b). Extending the
analysis from $z=0.3$ to $z=0$ (using SDSS data), Faber an colleagues find
$\Delta M^*\sim 1.3$ mag and a drop in $\phi^*$ by a factor of $\sim 4$
between $z=0$ and 1.1, but caution that much of these changes occur
between the $z\sim 0$ survey and their first bin (at $z=0.3$) at one end,
and in the last redshift bin at the other end, where the data are said
to be the weakest.

Both in COMBO-17 and DEEP2 the shape of the Schechter function for
the red-sequence galaxies is assumed constant with redshift. As such,
by construction this assumption virtually excludes down-sizing, for
which ubiquitous indications have emerged both at low as well as high
redshift. Indeed, as alluded in Kong et al. (2006) and
documented by Cimatti, Daddi \& Renzini (2006), COMBO-17 
and DEEP2
results can also be read in a different way. Figure 16 shows the
evolution of the rest frame $B$-band LF from COMBO-17,
%, DEEP2 and the
%Subaru/XMM Deep Survey (SXDS, Yamada et al. 2005), 
with the continuous line being the local LF for red-sequence galaxies
from Baldry et al. (2004).  The local LF has been shifted according to
the brightening derived from the empirical FP shift with redshift for
cluster ETGs (i.e., by $\Delta M_{\rm B}=-1.15\Delta z$, coming from
$\Delta(M/\lb)=-0.46z$ (van Dokkum \& Stanford 2003), taken as the
empirical template for passive evolution. From Figure 16 it is
apparent that the brightest part of the LF is fully consistent with
pure passive evolution of the most massive galaxies, whereas the
fainter part of the LF (below $\sim L^*$) is progressively depopulated
with increasing redshift, an effect that only in minor part could be
attributed to incompleteness.  Therefore, from these data it appears
that virtually all the most massive ETGs have already joined the red
sequence by $z\sim 1$, whereas less massive galaxies join it later.
This is what one would expect from the down-sizing scenario, as
exemplified e.g., in Figure 6, as if down-sizing was not limited to
stellar ages (stars in massive galaxies are older), but it would work
for the assembly itself, with massive galaxies being the first to be
assembled to their full size. Being more directly connected to the
evolution of dark matter halos, an apparent antihierarchical assembly
of galaxies may provide a more fundamental test of the $\Lambda$CDM
scenario than the mere down-sizing in star formation.

The slow evolution with redshift of the number density of
spectrum-selected bright ETGs was also one of the main results of the
K20 survey (Pozzetti et al. 2003), and more recently of the VLT VIMOS
Deep Survey (VVDS) where the rest-frame $B$-band LF of ETGs to $I<24$
is found to be broadly consistent with passive evolution up to $z\sim
1$, with the number density of bright ETGs decreasing by $\sim 40\%$
between $z=0.3$ and 1.1 (Zucca et al. 2006).

\begin{figure}[t]
%\vskip-0.8truecm
\centerline{\psfig{file=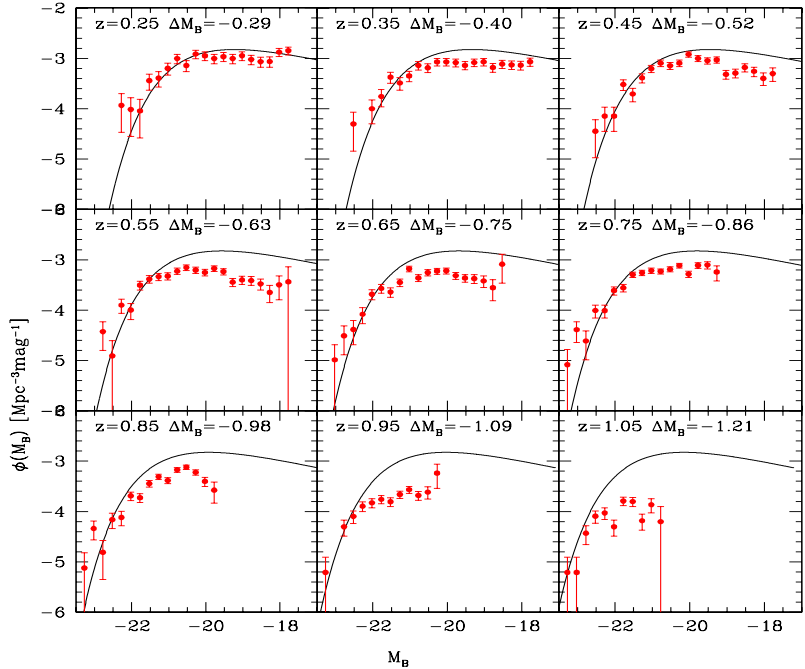,width=12.0cm,angle=0,height=10.cm}}
%\vskip-5truecm
%\plotone{feab.eps}
\caption{\piedi The evolution of the rest-frame $B$-band luminosity
function of early-type (red sequence) galaxies from COMBO-17 (Bell et
al. 2004) is compared to to the local luminosity function (solid line)
from Baldry et al. (2004). The local LF has been shifted in magnitude
as indicated in each panel, which corresponds to pure passive
evolution as empirically derived from the FP shift of cluster of
galaxies ($\Delta M/\lb = -0.46z$, van Dokkum \& Stanford 2003). Note
that there appears to be no number density evolution at the bright end
(i.e., for most massive galaxies), whereas at fainter magnitudes there
is a substantial decline with redshift of the number density of
ETGs. (From Cimatti, Daddi \& Renzini 2006.)
}
%\vskip-0.4 truecm
\end{figure}

Quite the same scenario, in which the most massive ETGs are already in
place at $z\sim 1$ while less massive ones appear later, emerges from
the study of the evolution of the stellar mass function for
morphologically-selected ETGs in the GOODS fields (Bundy, Ellis \&
Conselice 2005, Caputi et al. 2006, Franceschini et al. 2006), and 
especially from
the thorough re-analysis by Bundy et al. (2006) of the color-selected
ETGs from the DEEP2 survey. No such effect for
morphologically-selected ETGs in the GOODS-South field is mentioned in
a recent study by Ferreras et al. (2005), who report instead a steep
decrease in their number density with redshift. It seems fair to
conclude that to fully prove (or disprove) down-sizing in mass
assembly, and precisely quantify the effect, one needs to explore the
luminosity and mass functions to deeper limits than reached so far,
while extending the search to wider areas is needed to overcome cosmic
variance. An endeavour of this size requires an unprecedented amount
of observing time at virtually all major facilities, both in space and
on the ground. The COSMOS project covering 2 square degrees (Scoville
2005) is deliberately targeted to this end, providing public
multiwavelength data extending from X rays to radio wavelengths that
will allow astronomers to map the evolution of galaxies and AGNs in
their large scale structure context, and to derive photometric and
spectroscopic redshifts (Lilly 2005) to substantially fainter limits
than reached so far. Thus, it will be possible to directly assess the
interplay between AGN activity, star-formation onset and quenching,
merging, mass growth, and morphological differentiation over the
largest presently possible scale, hence promising substantial, perhaps
definitive progress in mapping the evolution of ETGs and their
progenitors.

It is worth emphasizing that even a low-level of ongoing star
formation can make galaxies drop out of our ETG samples, and that
theoretical models do not make solid predictions on when star
formation is going to cease in a dark matter halo. Therefore, in a
broader perspective, what is perhaps more fundamental than the
early-type/late-type distinction, is the stellar mass of galaxies and
the evolution of the galaxy mass function, irrespective of galaxy
type. Several ongoing studies are moving in this directions (e.g.,
Fontana et al. 2004, Drory et al. 2004, 2005, Bundy, Ellis, \&
Conselice 2005, Gabasch et al. 2005, Conselice, Blackburne, \&
Papovich 2005, etc.), but extending the discussion to the general mass
assembly of galaxies goes beyond the scope of this review.

\section{CATCHING ELLIPTICALS IN FORMATION}

Immediate precursors for cluster ETGs in the local universe were
first identified with a population of blue, star-forming galaxies
whose fraction in clusters increases very rapidly with redshift
(Butcher \& Oemler 1978, 1984, a trend known as the Butcher-Oemler
effect), and therefore most such blue, star-forming galaxies had to
subside to passive ETGs by $z\sim 0$.  Dressler el al. (1997) and
Fasano et al. (2000) showed that in clusters up to $z\sim 0.5$ the
fraction of true ellipticals is fairly constant, while the fraction of
(star-forming) spirals rapidly increases at the expense of the S0's.
This was interpreted as evidence that galaxies in clusters that were
star-forming spirals at $z\sim 0.5$, have changed their morphology to
become passively evolving S0's by $z=0$, a transformation that may
account for much of the Butcher-Oemler effect. In clusters in the same
redshift range, a sizable population of ETGs with strong Balmer
absorption lines was also identified by Dressler \& Gunn (1983), hence
called K+A galaxies after the appearance of their spectrum. These ETGs
were recognized as poststarburst galaxies, as further documented by
Dressler et al. (1999, 2004) and Poggianti et al.  (1999).

Thus, the metamorphosis from $z\sim 0.5$ to $z=0$ of a fraction of
cluster galaxies from star forming to passive is well documented by
these studies.  However, these changes seem to affect more spirals and
S0's than true ellipticals, and if the bulk of star formation in ETGs
took place at very high redshift, then we need to look further out to
catch them in formation. From the low-redshift studies we have learned
that ETGs are massive, highly clustered, and had to form the bulk of
their stars at $z\gsim 1.5-3$ (depending on mass and environment)
within a short time interval. Indeed, fast star formation is required
by the $\alpha$-overabundance, and by the mere high formation
redshift. At $z\sim 3$ the universe is only $\sim 2$ Gyr old, hence
forming $\sim 10^{11}\msun$ of stars in one object between $z=5$ and
$z=3$ ($\Delta t\sim 1$ Gyr) requires a star-formation rate (SFR) of
$\gsim 100\,\msy$.  Altogether, possible precursors of massive ETGs at
low $z$ could be searched among high-$z$ massive, highly clustered,
starburst galaxies with very high SFRs ($\gsim 100\,\msy$).  The rest
of this section is dedicated to mentioning the main results in
searching for such objects.

Lyman break galaxies (LBGs) were the first ubiquitous population of
galaxies to be identified at $z\sim 3$, and their SFRs often in excess
of $\sim 100\,\msy$, plus their relatively small size ($\re\sim 1-3$
kpc), made them natural precursors to local bulges and ETGs (e.g.,
Giavalisco, Steidel, \& Macchetto 1996, Steidel et al.  1996,
Giavalisco 2002, but see also Giavalisco et al. 1995 for an even
earlier attempt to identify a $z=3.4$ precursor to ETGs). Recently,
Adelberger et al. (2005) show that the 3D correlation length
($r_\circ$) of LBGs is such to match that of local lesser spheroids
($M\lsim 10^{11}\msun$) when the secular increase of $r_\circ$ is
taken into account. However, Adelberger et al. note that the most
massive and most rapidly star forming galaxies at high redshifts are
likely to be lost by the Lyman break selection. From Figure 2 we see
that local ETGs with $M>10^{11}\msun$ include almost 50\% of the mass
in this kind of galaxies, and $\sim 1/4$ of the total stellar mass at
$z=0$. Therefore, looking at the starforming precursors of the most
massive ETGs refers to a major component of the whole galaxy
population. Hence the search was not limited to the LBG technique.

Other obvious candidates are the ultraluminous infrared galaxies
(ULIRG, Sanders et al. 1988, Genzel \& Cesarsky 2000), detected in mm
or sub-mm surveys (e.g., Ivison et al. 1998, Lilly et al. 1999), a
class defined for their infrared luminosity ($8-1000\,\mu$m) exceeding
$\sim 10^{12}\lsun$, and whose typical SFRs ($\gsim 200 \,\msy$) well
qualify them for being precursors to massive spheroids, as does their
very high (stellar) density ensuring that they would ``land'' on the
fundamental plane as star formation subsides (Kormendy\& Sanders 1992,
Doyon et al. 1994). ULIRGs as ETG in formation are also advocated by
Genzel et al. (2001), who from the resolved kinematics for 12 of them
argue that typical ULIRGs are likely precursors to intermediate-mass
ETGs rather than to giant ellipticals. However, the internal
kinematics of one ULIRG at $z=2.8$ indicates a mass $\gsim 3\times
10^{11}\msun$ (Genzel et al. 2003) making it a likely precursor to a
very massive ETG.  Blain et al. (2004) note that
submillimeter-selected galaxies at $z=2-3$ appear to be more strongly
clustered than LBGs at the same redshifts, which makes them more
attractive candidates than LBGs for being the progenitors of the most
massive ETGs. Still, their space density falls short by a factor of
$\sim 10$ compared to passive EROs at $z\sim 1$, and they could be the
main precursors to EROs only if their duty cycle is very short.
Nevertheless, Chapman et al. (2005) argue that the sub-mm galaxies may
well be the dominant site of massive star formation at $z=2-3$, once
more making them excellent candidates for being ETG in formation.

The survey and characterization of sub-mm galaxies are currently
limited by the modest sensitivity and resolution of existing sub-mm
facilities, hence optical/near-IR selections are still the most
efficient way of identifying large samples of massive star-forming
galaxies at high redshift.  For the star-forming $BzK$-selected
objects, Daddi et al. (2004) estimate $<\!{\rm SFR}\!>\simeq
200\,\msy$, which is typical of ULIRGs, and most of them are clearly
mergers on ACS images.  Indeed, out of 131 non-AGN (i.e., non X-ray
emitter) star-forming BzKs with $K<20$ in the GOODS-North field
(Dickinson et al. in preparation), 82\% were individually detected
with Spitzer/MIPS at 24 $\mu$m (Daddi et al. 2005a). Moreover, by
stacking the fluxes of the 131 objects ($<\! z\!>=1.9$) from radio to
X-rays (i.e., VLA 1.4 GHz, SCUBA 850 and 450 $\mu$m, MIPS 24 $\mu$m,
IRAC 8-3.5 $\mu$m, near-IR and optical bands, and Chandra's 0.5-8 keV)
Daddi and colleagues showed that the resulting composite SED is an
excellent match to that of a template ULIRG with $L_{\rm IR}=1.7\times
10^{12}\lsun$ and $<\!{\rm SFR}\!>\simeq 250\,\msy$, in agreement with
the typical SFRs derived from the extinction-corrected UV flux. Two of
these BzKs have also been detected at 1.2 mm with MAMBO, implying a
SFR$\sim 1000\,\msy$ (Dannerbauer et al. 2006). So, the BzK selection
proves to be an excellent way of finding large numbers of ULIRGs at
high redshift, whose space density at $z\sim 2$ ($\sim 1-2\times
10^{-4}\mpcc$) is about three orders of magnitudes higher than the
local density of ULIRGs, and a factor of 2-3 higher than that at
$z=1$. Moreover, the number of star-forming BzKs with $M>10^{11}\msun$
is close to that of passive BzKs of similar mass, and added together
nearly match the space density of massive ETGs at $z=0$ (Kong et
al. 2006). Hence, it is tantalizing to conclude that as star formation
subsides in star-forming BzKs the number of passive ETGs will approach
their local density.  Finally, worth mentioning is that the majority
among samples of $J-K>2.3$ DRGs in the Extended HDF-South field (Webb
et al. 2006) and GOODS-South field (Papovich et al. 2006) have been
recently detected at 24 $\mu$m with Spitzer/MIPS, indicating that the
majority of DRGs are likely to be dusty starburst precursors to ETGs,
rather than having already turned into passive ETGs themselves. Moreover,
DRGs appear to be distributed over a very broad and nearly flat redshift 
distribution, from less than 1 to over $\sim 3.5$ (Reddy et al. 2006).

Moderate redshift precursors to local ETGs are not necessarily
star-forming. They may also be less massive ETGs that will merge by
$z=0$, an event now called ``dry merging''. The merger rate since
$z\sim 1.2$ has been estimated by Lin et al. (2004) as part of the
DEEP2 survey, concluding that only $\sim 9\%$ of present-day $M^*$
galaxies have undergone a major merger during the corresponding time
interval. However, Bell et al. (2006) searched for dry merger
candidates over the GEMS field, and based on 7 ETG-ETG pairs estimated
that each present-day ETG with $M_{\rm v}<-20.5$ has undergone 0.5--1
major dry merger since $z\sim 0.7$. This may be at variance with the
estimate based on the 3D two-point correlation function of local ETGs
in the overwhelming SDSS database. Indeed, each local ETG is found to have
less than 1\% probability per Gyr of merging with another ETG, hence
the dry merging rate appears to be ``lower, much lower than the rate
at which ETG-hosting DM halos merge with one another'' (Hogg 2006), at
least for $z\lsim 0.36$ (Masjedi et al. 2006).

The history of star formation in ETGs has been deduced from the
properties of their passively evolving stellar populations in low and
high redshifts galaxies, and we know that ETGs and bulges hold at
least $\sim 50\%$ of the stellar mass at $z=0$. Threfore, it is worth
addressing here one last issue, even if only in a cursory way: is such
an inferred history of star formation consistent with the direct
measurements of the star-formation density and stellar mass density at
high redshifts?  Based on the estimate that the bulk of stars in
spheroids formed at $z\gsim 3$, it has been suggested that at least
$\sim 30\%$ of all stars (and metals) have formed by $z=3$ (Renzini
1999). This appears to be at least a factor of $\sim 3$ higher than
the direct estimate based on the HDF-North, according to which only
$3\%-14\%$ of today's stars were in place by $z=3$ (Dickinson et
al. 2003). Data from HDF-South give the higher value $10\%-40\%$
(Fontana et al. 2003), most likely as a result of cosmic variance
affecting both HDF fields. Based on the $\sim 10$ times wider field of
the K20 survey, $\sim 30\%$ of the stellar mass appears to be in place
by $z\sim 2$ (Fontana et al. 2004), but the corrections for
incompleteness are large. Drory et al. (2005) find that over the $\sim
200$ arcmin$^2$ area of the combined GOODS-South and FORS Deep Field,
$\sim 50\%$, $\sim 25\%$, and at least $\sim 15\%$ of the mass in stars
is in place, respectively at $z=1$, 2, and 3. So, no gross discrepancy has 
emerged so far between the mass density at $z\sim 3$ as directly measured and
as estimated from the fossil evidence at lower redshift. The same
holds for the comparison between the star-formation
densities as a function of redshift, as inferred from the distribution
of stellar ages of ETGs on the one hand, and as directly measured by
observations on the other hand (e.g., Madau et al. 1996, Steidel et
al. 1999, Giavalisco et al. 2004b).  Errors on both sides are still as
large as a factor of 2 or 3, but this should rapidly improve
thanks to the deep and wide surveys currently under way.
\vfil\newpage
\section{EPILOGUE}

Almost thirty years ago Toomre (1977) remarked that star formation was
observed only in galaxy disks, and further that the final state of
merging spirals must be something resembling an elliptical galaxy.
Thus, merging spirals to form ellipticals at relatively low redshifts
became very popular, especially following the success of CDM theories 
in accounting for the growth of large scale structure from tiny initial
perturbations.

However, in the intervening three decades an impressive body of
evidence on galaxies at low as well as high redshift has accumulated,
that at least in part contradicts Toomre's assumption. While in the
local universe most of the star formation is indeed confined to disks; at
$z>1-1.5$ most of it appears to take place in starburst galaxies, such
as ULIRGs, whose space density is orders of magnitude higher than in
the local universe.  Moreover, $\sim 50\%$ of all stars seem to have
formed at $z\lsim 1$ (Dickinson et al. 2003), and to have occurred
mostly in disks (Hammer et al. 2005), whereas, if the scenario shown in
Figure 6 is basically correct, then the bulk of star formation in ETGs
took place at much higher redshift. At the risk of some
simplification, we can say that the era of ETG/spheroid/elliptical
formation was largely finished by $z\sim 1$ (if not before), just when the
major build up of disks was beginning (see also Papovich et al. 2005).

The evidence for the stellar populations in ETGs being old, and older
in massive galaxies than in less massive ones, has been known for over
ten years, along with the evidence for down-sizing and for the
anticorrelation of mass and SFR. Theoretical models based on the CDM
paradigm have recently incorporated these observational constraints,
and have been tuned to successfully reproduce the down-sizing effect
in star formation (e.g., De Lucia et al. 2006).  In a hierarchical
scenario, down-sizing in star formation is indeed natural.  Star
formation starts firsts in the highest density peaks, which in turn
are destined to become the most massive galaxies later on. But until
recently models predicted that star formation was continuing all the
way to low redshift, as cooling flows were left uncontrasted, thus
failing to even produce a red sequence. To get the old and dead
massive galaxies we see in nature, such cooling flows (and the
accompanying star formation) had to be suppressed in the models, which
is now generally accomplished by invoking strong AGN feedback, as
first incorporated in $\Lambda$CDM simulations by Granato et
al. (2001)\footnote{See also Ciotti et al. (1991) and Ciotti \&
Ostriker (1997) for early attempts to suppress cooling flows in ETGs,
either with Type Ia supernova feedback alone, or in combination with
AGN feedback.}. Yet, the AGN responsibility in switching off star
formation remains conjectural at this time, but we became aware that
galaxies and supermassive black holes co-evolve, which means we must
understand their formation as one and the same problem.

Baryon physics, including star formation, black hole formation and
their feedbacks, is highly nonlinear, and it is no surprise if
modelling of galaxy evolution relies heavily on many heuristic
algorithms, their parameterization, and trials and errors. Dark matter
physics, on the contrary, is extremely simple by comparison. Once DM halos are
set into motion, there is nothing preventing them from merging with
each other under the sole action of gravity, and growing bigger and
bigger ``galaxies'' in an up-sizing process. Thus, the 
vindication of the $\Lambda$CDM paradigm should be found in  observations 
demonstrating that the biggest, most massive galaxies are the first to
disappear when going to higher and higher redshifts.  This is indeed
what has not been seen yet, and actually there may be hints for the
contrary.  
\medskip\pn
ACKNOWLEDGMENTS
\medskip\pn I thank Ralf Bender, Andrea Cimatti, Emanuele Daddi, Mauro
Giavalisco, Laura Greggio, Silvia Pellegrini and Daniel Thomas for a
critical reading of the manuscript and for their valuable
suggestions. I am indebted to Mariangela Bernardi, Daniel Thomas, and
Sperello di Serego Alighieri, for having provided respectively Table
1, Figure 2, and Figure 13, specifically for this paper. Finally, I am
very grateful to my Annual Review tutoring editor John Kormendy for
his guidance, to Doug Beckner for the final set up of all the
figures, and to Roselyn Lowe-Webb for her patience in proof-editing the
manuscript.

%\vfill 

\end{document}